\documentclass[prb,preprint,showpacs,amsmath,amssymb,superscriptaddress]{revtex4}
\usepackage{graphicx}% Include figure files
\usepackage{dcolumn}% Align table columns on decimal point
\usepackage{bm}
\usepackage{natbib}
\usepackage{amssymb}
\usepackage{dcolumn}
\usepackage{eso-pic,everyshi,ifthen,calc}
\usepackage{pdfpages}

\begin{document}

\title{Establishing micromagnetic parameters of ferromagnetic semiconductor (Ga,Mn)As}
\author{P.~N\v{e}mec}
\affiliation{Faculty of Mathematics and Physics, Charles University in Prague, Ke Karlovu 3,
121 16 Prague 2, Czech Republic}
\author{V.~Nov\'ak}
\affiliation{Institute of Physics ASCR, v.v.i., Cukrovarnick\'a 10, 162 53 Praha 6, Czech Republic}
\author{N.~Tesa\v{r}ov\'a}
\affiliation{Faculty of Mathematics and Physics, Charles University in Prague, Ke Karlovu 3,
121 16 Prague 2, Czech Republic}

\author{E.~Rozkotov\'a}
\affiliation{Faculty of Mathematics and Physics, Charles University in Prague, Ke Karlovu 3,
121 16 Prague 2, Czech Republic}

\author{H.~Reichlov\'a}
\affiliation{Institute of Physics ASCR, v.v.i., Cukrovarnick\'a 10, 162 53 Praha 6, Czech Republic}
\affiliation{Faculty of Mathematics and Physics, Charles University in Prague, Ke Karlovu 3,
121 16 Prague 2, Czech Republic}

\author{D.~Butkovi\v{c}ov\'a}
\affiliation{Faculty of Mathematics and Physics, Charles University in Prague, Ke Karlovu 3, 121 16 Prague 2, Czech Republic}

\author{F.~Troj\'anek}
\affiliation{Faculty of Mathematics and Physics, Charles University in Prague, Ke Karlovu 3, 121 16 Prague 2, Czech Republic}

\author{K.~Olejn\'{\i}k}
\affiliation{Institute of Physics ASCR, v.v.i., Cukrovarnick\'a 10, 162 53 Praha 6, Czech Republic}

\author{P.~Mal\'y}
\affiliation{Faculty of Mathematics and Physics, Charles University in Prague, Ke Karlovu 3,
121 16 Prague 2, Czech Republic}

\author{R.~P.~Campion}
\affiliation{School of Physics and
Astronomy, University of Nottingham, Nottingham NG7 2RD, United Kingdom}

\author{B.~L.~Gallagher}
\affiliation{School of Physics and
Astronomy, University of Nottingham, Nottingham NG7 2RD, United Kingdom}

\author{Jairo~Sinova}
\affiliation{Department of Physics, Texas A\&M University, College
Station, TX 77843-4242, USA}
\affiliation{Institute of Physics ASCR, v.v.i., Cukrovarnick\'a 10, 162 53 Praha
6, Czech Republic}

\author{T.~Jungwirth}
\affiliation{Institute of Physics ASCR, v.v.i., Cukrovarnick\'a 10, 162 53 Praha 6, Czech Republic}
\affiliation{School of Physics and Astronomy, University of Nottingham, Nottingham NG7 2RD, United Kingdom}

\pacs{75.50.Pp,75.30.-m,75.70.Ak}
\date{\today}
\maketitle

{\bf 
(Ga,Mn)As is at the forefront of research exploring the synergy of magnetism with the physics and technology of semiconductors, and has led to discoveries of new spin-dependent phenomena and functionalities applicable to a wide range of material systems. Its recognition and utility as an ideal model material for spintronics research has been undermined by the large scatter in reported semiconducting doping trends and micromagnetic parameters. In this paper we establish these basic material characteristics by individually optimizing the highly non-equilibrium synthesis  for each Mn-doping level and by simultaneously determining  all micromagnetic parameters from one set of magneto-optical pump-and-probe measurements. Our (Ga,Mn)As thin-film epilayers, spannig the wide range of accessible dopings, have sharp thermodynamic Curie point singularities typical of uniform magnetic systems. The materials show systematic trends of increasing magnetization, carrier density, and  Curie temperature (reaching 188~K) with increasing doping, and monotonous doping dependence of the Gilbert damping constant of $\sim {\bf 0.1-0.01}$ and the spin stiffness of $\sim {\bf 2-3}$~meV\,nm$^2$. These results render (Ga,Mn)As well controlled degenerate semiconductor with basic magnetic characteristics comparable to common band ferromagnets.} 

Under equilibrium growth  conditions the incorporation of magnetic Mn ions into III-As semiconductor crystals is limited to approximately  0.1\%.  To circumvent  the solubility problem  a non-equilibrium, low-temperature molecular-beam-epitaxy (LT-MBE) technique was employed which led to first successful growths of (In,Mn)As and (Ga,Mn)As ternary alloys with more than 1\% Mn and to the discovery of ferromagnetism in these materials.\cite{Ohno:1992_a,Munekata:1993_a,Ohno:1996_a,Hayashi:1997_a,VanEsch:1997_a,Ohno:1998_a} The compounds qualify as ferromagnetic semiconductors to the extent that their magnetic properties can be altered by the usual semiconductor electronics engineering variables, such as doping, electric fields,\cite{Ohno:2000_a,Chiba:2003_a,Chiba:2008_a,Olejnik:2008_a,Owen:2008_a,Stolichnov:2008_a} or light.\cite{Munekata:1997_a,Koshihara:1997_a,Ohno:1999_b,Oiwa:2005_a,Wang:2006_b,Takechi:2007_a,Qi:2007_a,Qi:2009_a,Rozkotova:2008_a,Rozkotova:2008_b,Hashimoto:2008_a,Hashimoto:2008_b,Kobayashi:2010,Nemec:2012_a,Tesarova:2012_a} By exploiting the large spin polarization of carriers and low saturation moment in (Ga,Mn)As and building on the well established heterostructure growth and microfabrication techniques in III-V semiconductors, (Ga,Mn)As has been extensively used for spintronics research of direct and inverse magneto-transport phenomena.\cite{Tanaka:2001_a,Chiba:2004_a,Saito:2005_a,Mattana:2005_a,Yamanouchi:2004_a,Yamanouchi:2006_a,Wunderlich:2007_c,Adam:2009_a,Wang:2010_a,Curiale:2012_a} Besides the more conventional spintronic effects based on Mott's two-spin-channel model of conduction in ferromagnets, (Ga,Mn)As has become particularly fruitful for exploring the second, more physically intriguing spintronics paradigm based on Dirac's spin-orbit coupling.\cite{Wunderlich:2007_c,Wenisch:2007_a,Rushforth:2008_a,Overby:2008_a,Goennenwein:2008_a,Gould:2004_a,Wunderlich:2006_a,Ciccarelli:2012_a,Chernyshov:2009_a,Fang:2010_a}

The apparent potential of (Ga,Mn)As to become the test-bed model material for many lines of spintronics research has been hindered by the large scatter in reported semiconducting doping trends and micromagnetic parameters. Our strategy to tackle this problem begins from the synthesis of a set of (Ga,Mn)As materials spanning a wide range of Mn dopings. Because of the highly non-equilibrium nature of the heavily-doped ferromagnetic (Ga,Mn)As, the growth and post-growth annealing procedures have to be individually optimized for each Mn-doping level in order to obtain films which are as close as possible to idealized uniform (Ga,Mn)As mixed crystals with the minimal density of compensating and other unintentional defects. An extensive set of characterization measurements has to accompany the synthesis to guarantee that the materials show systematic doping trends; monitoring the thermodynamic Curie point singularities is essential for assuring the uniformity and high magnetic quality of the materials.\cite{Olejnik:2008_a,Novak:2008_a,Wang:2008_e,Jungwirth:2010_b} When omitting the above procedures,\cite{Dobrowolska:2012_a}  extrinsic impurities and sample inhomogeneities can yield non-systematic doping trends and conceal the intrinsic magnetic properties of (Ga,Mn)As. 

\begin{figure}[h!]
%\vspace*{-0.2cm}
\hspace{0cm}\includegraphics[width=.7\columnwidth,angle=0]{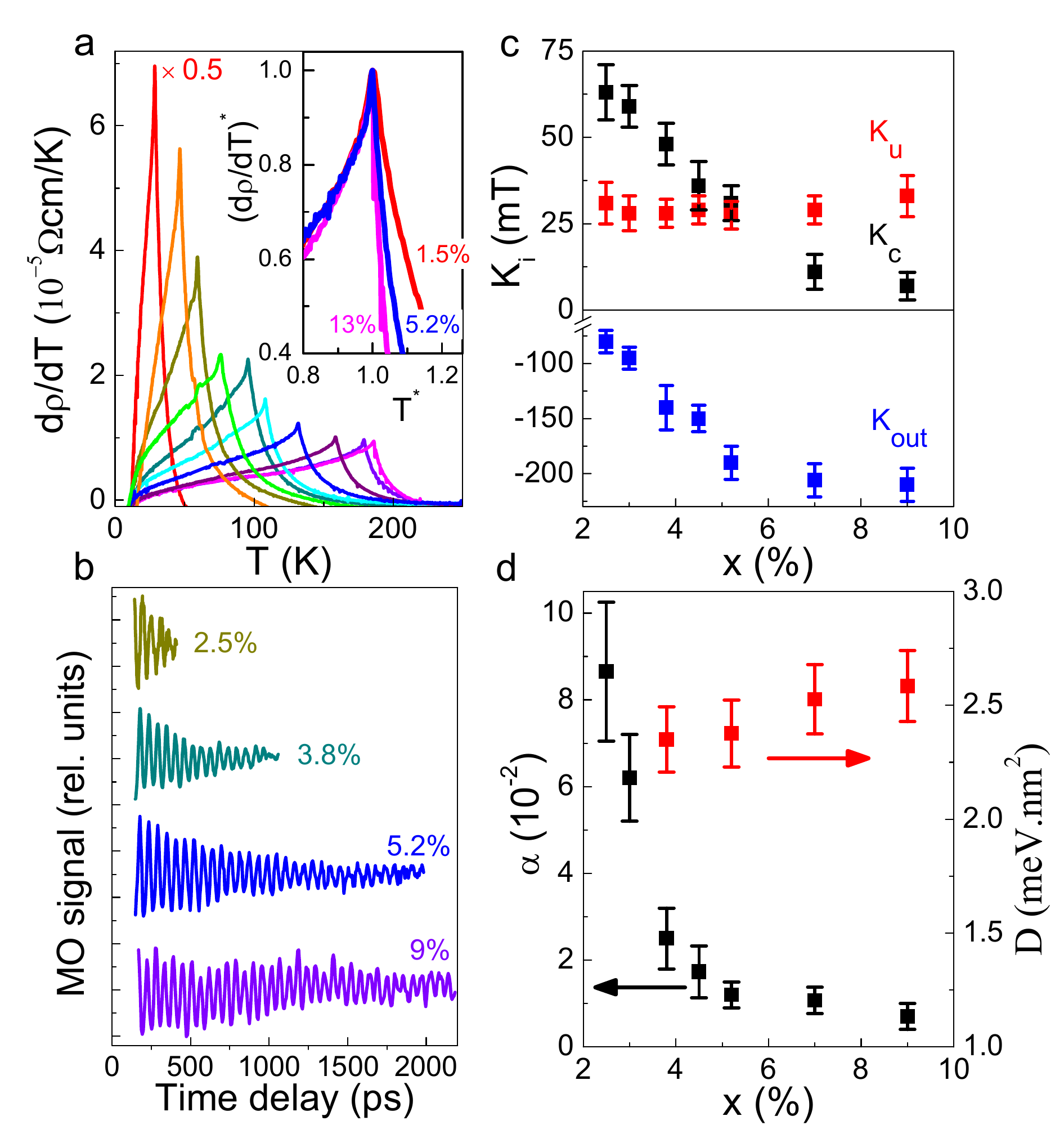}
%\vspace*{-0.5cm}
\caption{{\bf Micromagnetic parameters of optimized epilayers of ferromagnetic (Ga,Mn)As.} {\bf a}, Examples of sharp Curie point singularities in the temperature derivative of the resistivity in the series of optimized ferromagnetic (Ga,Mn)As epilayers with metallic conduction; $T_c$ monotonously increases with increasing nominal Mn doping between 1.5 and 13\%. Inset shows $d\rho/dT$  normalized to its peak value  with the temperature axis normalized to  $T_c$. {\bf b,} Examples of oscillatory parts of MO signals measured in 18 nm thick (Ga,Mn)As epilayers with the depicted nominal Mn doping for external magnetic field $\mu_0H_{ext} = 400$~mT applied along the [010] crystallographic direction; the curves are normalized and vertically off-set for clarity. {\bf c,} Dependence of anisotropy constants on nominal Mn doping. {\bf d,} Dependence of the Gilbert damping constant $\alpha$ and the spin stiffness constant $D$ on nominal Mn doping.}
\label{fig1}
\end{figure}

The focus of the work presented in this paper is on the systematic study of the  Gilbert damping and spin stiffness constants which, together with magnetic anisotropy fields, represent the basic micromagnetic parameters of a ferromagnet. A more than an order of magnitude experimental scatter and a lack of any clear trend as a function of Mn-doping can be found in the literature for  the Gilbert damping and spin-stiffness constants.\cite{Sinova:2004_b,Khazen:2008_a,Rappoport:2004_a,Zhou:2007_a,Liu:2007_e,Bihler:2009_a,Potashnik:2002_a,Gourdon:2007_a,Wang:2007_f,Werpachowska:2010_a} (See Supplementary information for a detailed discussion of previous experimental works.) This reflects partly the issues related to the control and reproducibility of the synthesis of (Ga,Mn)As and partly the difficulty with applying common magnetic characterization techniques, such as neutron scattering, to the thin-film dilute-moment (Ga,Mn)As samples.  
Hand-in-hand with the optimization of the material synthesis we have developed experimental capabilities based on the magneto-optical (MO) pump-and-probe method which allow us to simultaneously determine the magnetic anisotropy, Gilbert damping, and spin stiffness constants from one consistent set of measured data. Our results are summarized in Fig.~1. The Curie point singularity in the temperature derivative of the resistivity $d\rho/dT$ measured throughout the series of optimized ferromagnetic (Ga,Mn)As samples with metallic conduction is shown in Fig.~1a. The data span the nominal doping range from $x\approx$1.5 to 13\% and corresponding Curie temperatures from $T_c=29$ to 188~K, and illustrate the high quality of all the epilayers within the series. Examples of the measured magnetization precession signals by the MO pump-and-probe method are shown in Fig.~1b. From these time-dependent magnetization measurements we obtained the magnetic anisotropy constants $K_i$, Gilbert damping constant $\alpha$, and spin stiffness constant $D$  which are summarized in Figs.1c,d. We now proceed to the detail discussion of our experimental techniques and the discussion of the measured results in the context of physics of degenerate semiconductors and band ferromagnets.

{\bf\em Optimization of the (Ga,Mn)As synthesis.} Our (Ga,Mn)As layers were grown at the growth rate of approximately 0.2 monolayers/second. The Mn flux, and hence the nominal Mn doping $x$, was determined by measuring the ratio of the beam equivalent pressures (BEP) of Mn and Ga sources before each growth. The Mn content was cross-checked by secondary ion mass spectroscopy (SIMS) and by comparing the growth rates of GaAs and (Ga,Mn)As measured by the oscillations of the reflection high-energy electron diffraction (RHEED).

\begin{figure}[h!]
%\vspace*{-0.2cm}
\hspace{0cm}\includegraphics[width=.7\columnwidth,angle=0]{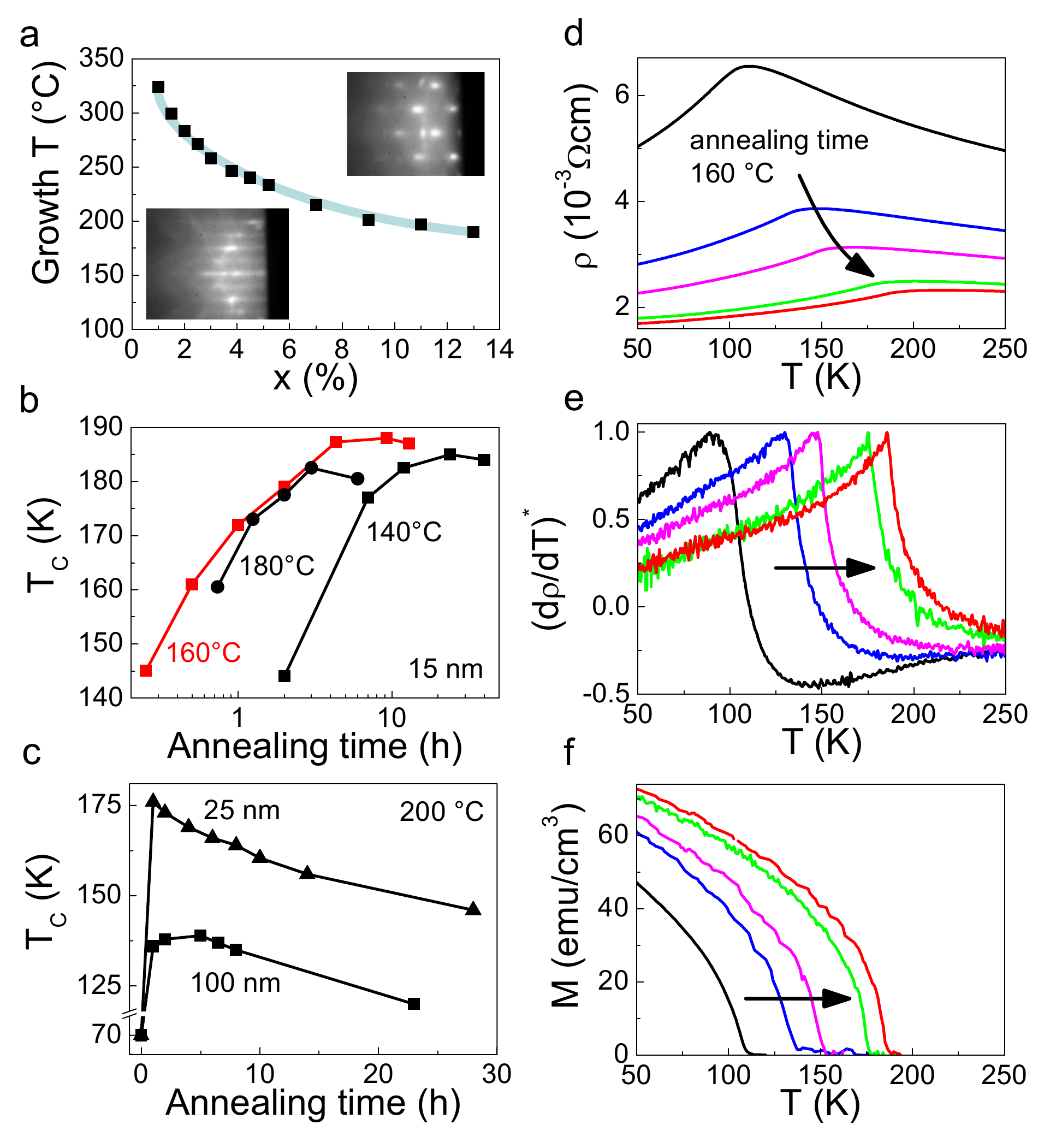}
%\vspace*{-0.5cm}
\caption{{\bf Optimization of the (Ga,Mn)As synthesis.} {\bf a}, Optimal growth temperature as a function of the nominal Mn doping. Insets show  examples of RHEED images of the 2D growth at 210$^\circ$C (lower inset) and 3D growth at 225$^\circ$C (upper inset) of the 7\% Mn-doped (Ga,Mn)As. 
{\bf b}, Dependence of the Curie temperature on the annealing time for three different annealing temperatures in a 15~nm thick (Ga,Mn)As epilayer with 13\% nominal Mn doping. 
{\bf c}, Dependence of the Curie temperature on the annealing time  for the annealing temperature of 200$^\circ$C in a 100~nm thick (Ga,Mn)As epilayer with 13\% nominal Mn doping, and in the same epilayer thinned down to 25~nm by wet etching.
{\bf d--f}, Temperature dependencies of resistivity $\rho$, {\bf d}, temperature derivative of the resistivity $d\rho/dT$, {\bf e}, and remnant magnetization $M$, {\bf f}, in a 20~nm thick (Ga,Mn)As epilayer with 13\% nominal Mn doping at successive annealing times at the optimal annealing temperature of 160$^\circ$C for this doping.}
\label{fig2}
\end{figure}

There are two critical growth parameters of (Ga,Mn)As: the substrate temperature, and the As-to-(Ga+Mn) flux ratio. At the typical temperatures of $\sim200^\circ$C neither an optical pyrometer nor a radiatively coupled temperature sensor are applicable. Instead, we used the GaAs band-edge spectrometer to measure the substrate temperature and the predictive substrate heater control to stabilize the temperature during the growth. For a given As:(Ga+Mn) ratio the substrate temperature fully determines the growth  regime: the growth proceeds two-dimensionally at low temperatures, and turns irreversibly into the 3D growth mode when a critical temperature is exceeded. The scatter of the critical substrate temperature for given $x$ and As:(Ga+Mn) ratio is remarkably small, typically less than 2$^\circ$C. In excess As flux the 2D/3D transition occurs at higher temperature. The highest quality samples are grown  in a narrow window of the 1:1 stoichiometric As:(Ga+Mn) ratio and at the substrate temperature approaching as close as possible from below the 2D/3D critical temperature for given $x$. The As:(Ga+Mn) ratio was adjusted by the As-cell valve, and calibrated using the As-controlled RHEED oscillations. In insets of Fig.~2a we show examples of RHEED patterns for the  $x=7\%$ nominally doped (Ga,Mn)As material grown at stoichiometric 1:1 ratio of As:(Ga+Mn) for substrate temperature of 225~K which is above the 3D/2D boundary and 210~K which is below the boundary. The optimal growth temperature for this doping is  215~K. In the main panel of Fig.~2a we plot the optimal growth temperature as a function of nominal Mn doping, showing the rapidly decreasing growth temperature trend.  

The next important factor determining the quality of the resulting (Ga,Mn)As materials are post-growth annealing conditions. In Fig.~2b we show the dependence of the Curie temperature $T_c$ on the annealing time for three different annealing temperatures for the record $T_c=188$~K sample with nominal 13\% Mn doping and film thickness 15~nm. These curves illustrate the common trend in annealing (at temperatures close to the growth temperature) suggesting the presence of competing mechanisms. One mechanism yields the increase of $T_c$ and is ascribed in a number of reports to the removal of charge and moment compensating interstitial Mn impurities (see e.g. the detailed annealing study in Ref.~\onlinecite{Olejnik:2008_a}). The removal is slowed down by the growth of an oxide surface layer during annealing\cite{Olejnik:2008_a} and an additional mechanism  can eventually yield reduction of  $T_c$ after sufficiently long annealing times, depending on the annealing temperature. The origin of this detrimental mechanism may be in Mn clustering or in the competition between the non-equilibrium (Ga,Mn)As phase and  the equilibrium MnAs second phase.   Because of the  competing mechanisms, the absolutely highest Curie temperature for the given nominal doping is achieved at intermediate annealing temperature and time, as illustrated in Fig.~2b. 

The remaining critical parameter of the synthesis is the epilayer thickness. For a given nominal doping, the highest attainable $T_c$ is reached only in thin films, typically thinner than $\sim$50~nm. In Fig.~2c we illustrate the importance of the film thickness for obtaining high quality (Ga,Mn)As materials.  A 100~nm thick film is grown with nominal 13\% doping and, unlike the thin record $T_c$ film discussed above, here the maximum $T_c$ achieved by annealing is only about 140~K. However, if the same film is thinned down (to e.g. 25~nm) by wet etching and annealed at the same conditions, the achieved Curie temperatures are significantly higher. 

An increase of $T_c$ is not the only parameter followed to ascertain that a sample is of high quality. A key characterization tool are the thermodynamic Curie point singularities.\cite{Novak:2008_a} This is illustrated in Figs.~2d-f where we compare resistivity and magnetization  measured at increasing time steps during the optimizing annealing procedure. The development of sharply vanishing magnetization $M(T)$ at $T_c$ and the onset of the singularity in $d\rho/dT$ are well correlated with increasing $T_c$ and conductivity within the annealing sequence.

\begin{figure}[h!]
%\vspace*{-0.2cm}
\hspace{0cm}\includegraphics[width=.7\columnwidth,angle=0]{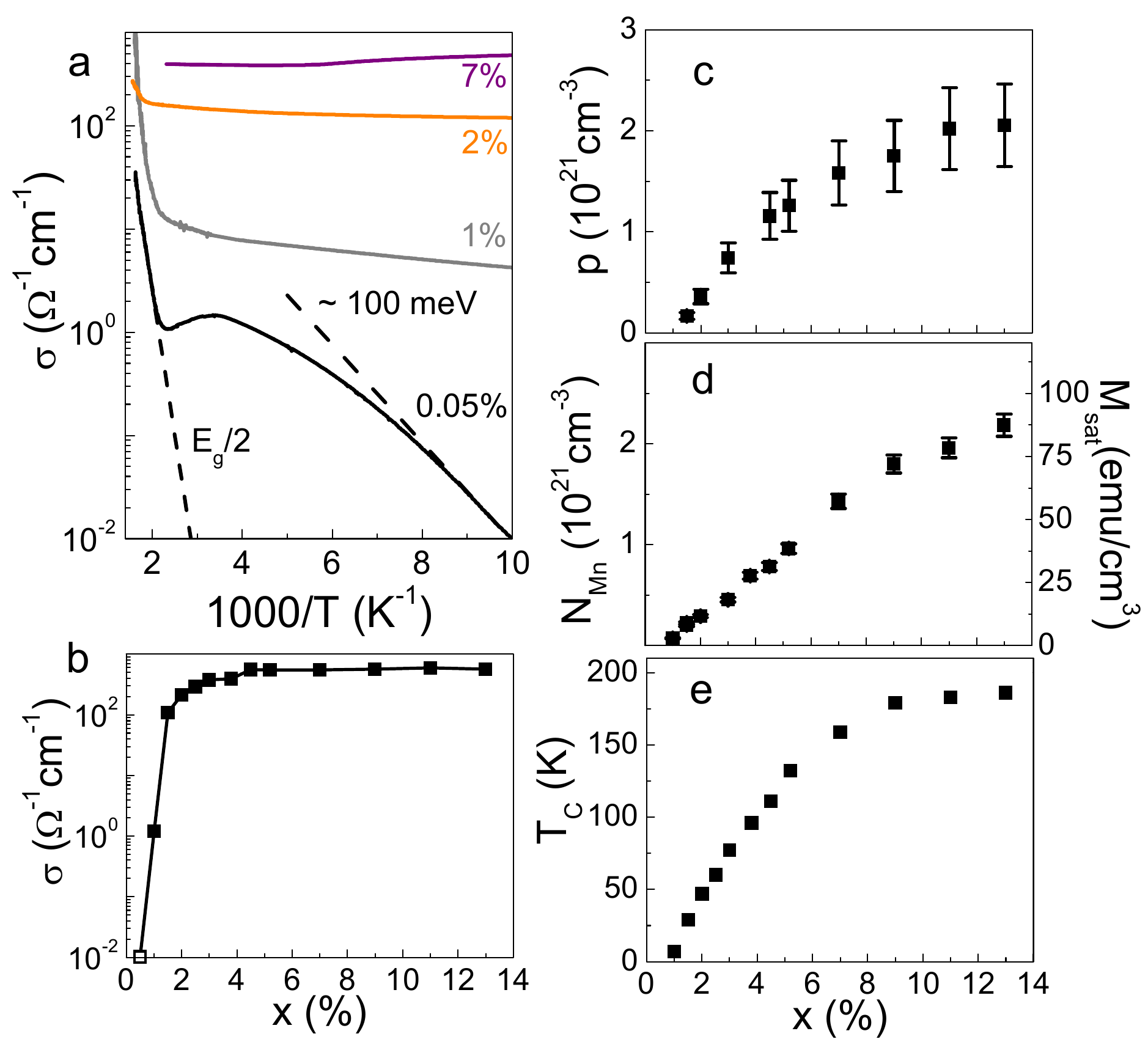}
%\vspace*{-0.5cm}
\caption{{\bf Doping trends in the series of optimized (Ga,Mn)As epilayers.} {\bf a}, Temperature dependence of the conductivity $\sigma(T)$ of optimized (Ga,Mn)As epilayers with depicted nominal Mn doping. Dashed lines indicate the activated parts of $\sigma(T)$ of the insulating paramagnetic (Ga,Mn)As with 0.05\% Mn doping, corresponding to the Mn acceptor level and the band gap, respectively.
{\bf b-e}, Conductivity, {\bf b}, hole density, {\bf c}, saturation magnetization and corresponding Mn moment density, {\bf d}, and Curie temperature, {\bf e}, as a function of the nominal Mn doping in the series of optimized (Ga,Mn)As epilayers. 
}
\label{fig3}
\end{figure}

After finding the optimal growth and post-growth conditions for each individual nominal doping we obtained a series of samples spanning the wide range of Mn dopings. The samples can be divided into several groups: at nominal dopings below $\sim0.1\%$ the (Ga,Mn)As materials are paramagnetic,  strongly insulating, showing signatures of the activated transport corresponding to valence band -- impurity band transitions at intermediate temperatures, and valence band -- conduction band transitions at high temperatures (see Fig.~3a).\cite{Jungwirth:2007_a,Masek:2010_a} For higher nominal dopings, $0.5 \lesssim x \lesssim 1.5\%$, no clear signatures of activation from the valence band to the impurity band are seen in the dc transport, confirming that the bands start to overlap and mix, yet the materials remain insulating.\cite{Jungwirth:2007_a,Masek:2010_a} At $x\approx1.5\%$, the low-temperature conductivity of the film increases abruptly by several orders of magnitude (see Fig.~3b), and the system turns into a degenerate semiconductor.\cite{Jungwirth:2007_a,Masek:2010_a} The onset of ferromagnetism occurs already on the insulating side of the transition at $x\approx 1\%$ and the Curie temperature then steadily increases with increasing nominal Mn doping up to $\approx13\%$. The hole concentration $p$ can be measured by the slope of the Hall curve at high fields (see Supplementary information) with an error bar due to the multi-band nature estimated to $\sim 20\%$.\cite{Jungwirth:2005_b} Within this uncertainty, the overall trend shows increasing $p$ with increasing doping in the optimized materials, as shown in Fig.~3c. Similarly, the saturation moment and $T_c$ steadily increase with increasing nominal doping up to $x\approx 13\%$, as shown in Figs.~3d,e. Assuming 4.5$\mu_B$ per Mn atom \cite{Jungwirth:2005_a} we can estimate the density $N_{Mn}$ of uncompesated Mn$_{\rm Ga}$ moments from the magnetization data (see left y-axis in Fig.~3d). An important conclusion can be drawn when comparing this estimate with the hole density estimated from the Hall resistance. Since there is no apparent deficit of $p$ compared to $N_{Mn}$, and since the interstitial Mn impurity compensates one local moment but two holes we conclude that interstitial Mn is completely (within the experimental scatter) removed in our optimally annealed epilayers. 
Hence, our series of optimized (Ga,Mn)As materials have reproducible characteristics, showing an overall trend of increasing saturation moment with increasing $x$ , increasing $T_c$ (reaching 188~K), and increasing hole density. The materials have no measurable
charge or moment compensation of the substitutional Mn$_{\rm Ga}$ impurities and have a large degree of uniformity reflected by sharp Curie point singularities. 

{\bf\em Determination of the micromagnetic parameters.} We now proceed to the determination of the magnetic anisotropy, Gilbert damping, and spin stiffness  constants of our (Ga,Mn)As epilayers from the MO time-resolved measurements of the magnetization precession. In the MO pump-and-probe experiments, we used a femtosecond titan sapphire laser that was spectrally tuned to 1.64 eV, i.e., above the band gap of GaAs. The possibility to excite and detect precession of ferromagnetic Mn moments in (Ga,Mn)As by this method has been extensively discussed in previous MO studies.\cite{Oiwa:2005_a,Wang:2006_b,Takechi:2007_a,Qi:2007_a,Qi:2009_a,Rozkotova:2008_a,Rozkotova:2008_b,Hashimoto:2008_a,Hashimoto:2008_b,Kobayashi:2010,Nemec:2012_a,Tesarova:2012_a} All  experiments presented below were preformed at temperature of approximately 15 K in reflection geometry. External magnetic fields up to 550 mT were applied in the [010] and [110] crystallographic directions. The intensity of the pump pulse was $\sim30-40$~$\mu$Jcm$^{-2}$, with the pump to probe intensity ratio $\sim20:1-10:1$. The penetration depth of the laser beam ($\sim600$~nm) safely exceeds the thickness of the studied (Ga,Mn)As epilayers.

The anisotropy constants, shown in Fig.~1c, where obtained combining three complementary measurements. In the first experiment we measured the external magnetic field $H_{ext}$ dependence of the precession frequency $f$ of the time resolved MO signal. In the studied (Ga,Mn)As/GaAs epilayers, the internal magnetic anisotropy fields are dominated by three components. The out-of-plane component $K_{out}$ is a sum of the thin-film shape anisotropy and the magnetocrystalline anisotropy due to the compressive growth strain in (Ga,Mn)As. The cubic magnetocrystalline anisotropy $K_{c}$ reflects the zinc-blende crystal structure of the host semiconductor. The additional uniaxial anisotropy component along the in-plane diagonal $K_{u}$ is not associated with any measurable macroscopic strain in the epilayer and is likely of extrinsic origin. The precession frequency is given by,
\begin{eqnarray}
f&=&\frac{g\mu_B}{h}\sqrt{\left(H_{ext}\cos(\varphi-\varphi_H)-2K_{out}+K_c(3+\cos4\varphi)/2+2K_{u}\sin^2(\varphi-\pi/4)+\Delta H_n\right)}\nonumber\\&\times&\sqrt{\left(H_{ext}\cos(\varphi-\varphi_H)+2K_c\cos4\varphi-2K_{u}\sin2\varphi+\Delta H_n\right)}\,,
\label{omega}
\end{eqnarray}
where $g$ is the Land\'e g-factor of Mn moments, $\mu_B$ the Bohr magneton, $\varphi$ and $\varphi_H$ are the in-plane magnetization and external magnetic field angles measured from the [100] crystal axis, and $\Delta H_n$ is the shift of the resonant field for the higher index $n$ spin wave modes with respect to the $n=0$ uniform precession mode. In order to uniquely determine the anisotropy constants, the field-dependent precession frequency measurements were complemented  by MO experiments with variable polarization angle of the probe beam. The latter measurements allow us to precisely determine the angle of the equilibrium easy axis of the magnetization (see Supplementary information).\cite{Nemec:2012_a,Tesarova:2012_a} Finally, we confirmed the consistency of the obtained anisotropy constants by performing static measurements of magnetization hysteresis loops by the superconducting quantum interference device (SQUID). Results for ferromagnetic materials from our series of optimized (Ga,Mn)As   epilayers are summarized in Fig.~1c. Note that the values of $K_{out}$ and $K_c$ for the given Mn-doping are well reproducible in materials whose synthesis yields the same optimized values of the basic structural, magnetic and transport properties. For the $K_{u}$ constant, variations in the width of the optimized thin (Ga,Mn)As films  or of other otherwise insignificant changes of the growth or annealing conditions may yield sizable changes of $K_{u}$. This confirms the presumed subtle extrinsic nature of this magnetic anisotropy component. 

The sign of $K_{out}$ implies that all studied (Ga,Mn)As/GaAs materials are in-plane ferromagnets. The competing magnitudes of $K_c$ and $K_{u}$ and the different doping trends of these two in-plane magnetic anisotropy constants (see Fig.~1c) are therefore crucial for the micromagnetics of the materials. The biaxial anisotropy $K_c$ dominates at very low dopings and the easy axis aligns with the main crystal axis [100] or [010]. At intermediate dopings, the uniaxial anisotropy $K_{u}$ is still weaker but comparable in magnitude to $K_c$. In these samples the two equilibrium easy-axes are tilted towards the [1$\bar1$0] direction and their angle is sensitive to small changes of external parameters such as temperature.  This allows for exciting the magnetization precession by laser pulses in the pump-and-probe MO experiments. At very high dopings, the uniaxial anisotropy dominates and the system has one strong easy-axis along the [1$\bar1$0] in-plane diagonal. In the low-doped and high-doped samples with very stable easy-axes aligned with one of the main crystal directions the dynamical MO experiments become unfeasible.  

The Gilbert damping constant $\alpha$, shown in Fig.~1d, is obtained by fitting the measured dynamical MO signal to Landau-Lifshitz-Gilbert (LLG) equations using the experimentally obtained magnetic anisotropy constants. The high accuracy of the LLG fits is demonstrated  in Figs.~4a,b on data measured in a $x=5.2$\% doped sample. The obtained dependence of $\alpha$ on the external magnetic field applied along the [010] and [110] directions is shown in Fig.~4c. At smaller fields, $\alpha$ is not constant and shows a strong anisotropy with respect to the field angle. When plotted as a function of frequency, however, the dependence on the field-angle disappears, as shown in Fig.~4d. Analogous results are obtained for the entire series of the optimized materials. We can therefore conclude that the apparent anisotropy of $\alpha$ can in our materials be ascribed fully to the field-angle dependence via the precession frequency. In all our studied materials, the frequency-independent Gilbert damping constant is isotropic and can be accurately determined from MO data with precession frequencies $f\gtrsim15$~GHz.

\begin{figure}[h!]
%\vspace*{-0.2cm}
\hspace{0cm}\includegraphics[width=.7\columnwidth,angle=0]{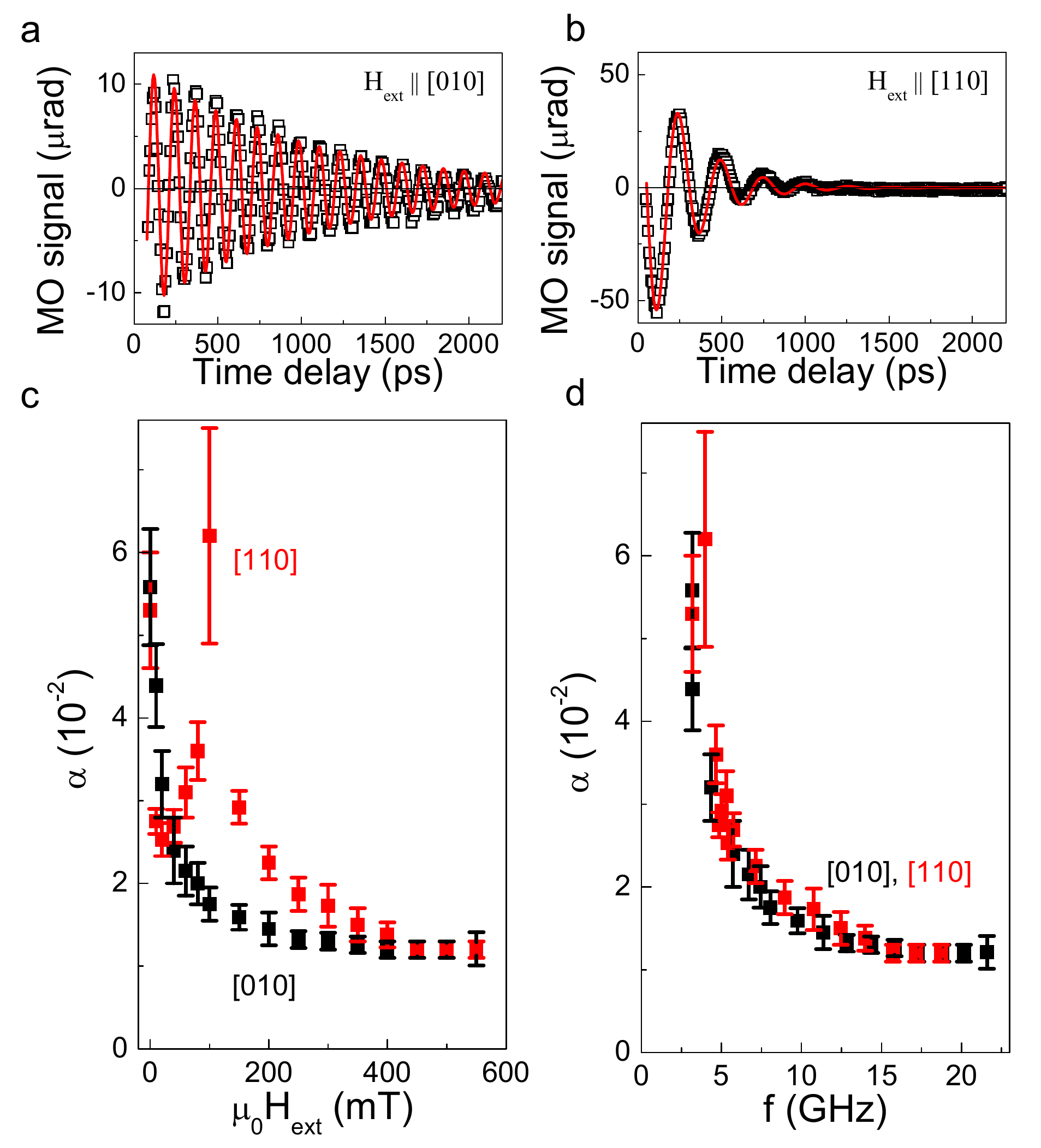}
%\vspace*{-0.5cm}
\caption{
{\bf Determination of the Gilbert damping constant from MO experiments.} {\bf a,b,} Oscillatory part of the MO signal (points) measured in a 18~nm thick epilayer with 5.2\% nominal Mn doping for external magnetic field $\mu_0 H_{ext} = 100$~mT applied along the crystallographic directions [010] and [110]; lines are fits by the LLG equation. {\bf c,} Dependence of the Gilbert damping on external magnetic field applied along the [010] and [110] crystallographic directions. {\bf d,} Dependence of the Gilbert damping on the precession frequency.
}
\label{fig4}
\end{figure}

We point out that in ferromagnetic resonance (FMR) experiments, the measurement frequency was limited to two values, $f=9$ and 35~GHz which even in the optimized (Ga,Mn)As materials is not sufficient to reliably separate the intrinsic Gilbert damping constant from the inhomogeneous broadening of the FMR line-width. The dynamical MO measurements, on the other hand, span a large enough range of frequencies and allow us to extract a consistent set of frequency-independent values of $\alpha$ for our series of optimized ferromagnetic (Ga,Mn)As materials.  We find a systematic doping trend across the series in which the Gilbert constant decreases from $\sim 0.1$ to $0.01$ when the nominal Mn doping increases from $\sim 2\%$ to $5\%$ and then remains nearly constant (see Fig.~1d). The magnitudes of $\alpha$ and the doping dependence are consistent with Gilbert damping constants in conventional transition metal ferromagnets. In metals, $\alpha$ typically increases with increasing resistivity and is enhanced in alloys with enhanced spin-orbit coupling.\cite{Ingvarsson:2002_a,Rantschler:2007_a,Gilmore:2008_a} Similarly, in our measurements in (Ga,Mn)As, the increase of $\alpha$ correlates with a sizable increase of the resistivity in the lower Mn-doped samples. Also, the spin-orbit coupling effects tend to be stronger in the lower doped samples with lower filling of the valence bands and with the carriers closer to the metal-insulator transition.\cite{Jungwirth:2006_a}  Theory ascribing magnetization relaxation to the kinetic-exchange coupling of Mn moments with holes residing in the disordered, exchange-split, and spin-orbit-coupled valence band of (Ga,Mn)As yields a comparable range of values of $\alpha$ as observed in our measurements.\cite{Sinova:2004_b}

Similar to the Gilbert constant, there has been a large scatter\cite{Werpachowska:2010_a} in previous reports of experimental values of the spin-stiffness in (Ga,Mn)As inferred from FMR,\cite{Rappoport:2004_a,Zhou:2007_a,Liu:2007_e,Bihler:2009_a} magneto-optical studies,\cite{Wang:2007_f}  and from complementary static magnetization and domain structure measurements.\cite{Potashnik:2002_a,Gourdon:2007_a} We attribute the lack of  a consistent picture obtained from these measurements to sample inhomogeneities and extrinsic defects in the studied (Ga,Mn)As epilayers with thicknesses typically exceeding 100~nm and to experimental data which allowed only an indirect extraction of the spin stiffness constant. The MO pump-and-probe technique utilized in our work allows in principle for the direct measurement of the spin stiffness, however, one has to find the rather delicate balance between thin enough epilayers to avoid sample inhomogeneity and thick enough films allowing to observe the higher-index  Kittel spin-wave modes\cite{Kittel:1958_a} of a uniform thin-film ferromagnet. For these modes, the spin-stiffness parameter $D$ is directly obtained from the measured resonant fields,
\begin{equation}
\Delta H_n\equiv H_0-H_n=D\frac{n^2}{L^2}\frac{\pi^2}{g\mu_B}\,,
\label{Hn}
\end{equation}
where $L$ is the thickness of the ferromagnetic film. The MO pump-and-probe technique has the key advantage here that, unlike FMR, it is not limited to odd index spin wave modes.\cite{Kittel:1958_a} The ability to excite and detect the $n=0$, 1, and 2 resonances is essential for the observation of the Kittel modes in our optimized  (Ga,Mn)As epilayers whose thickness is limited to $\sim50$~nm.

In Fig. 5a we show an example of the time dependent MO signal measured in a 48~nm thick optimized epilayer with 7\% nominal Mn doping. Three spin wave resonances (SWRs) are identified in the sample with frequencies $f_0$,  $f_1$, and $f_2$, as shown in Figs.~5b,c. The association of these SWRs with the Kittel modes, described by Eq.~(\ref{Hn}), is based on experiments shown in Figs.~5b-e.  In Fig. 5c we plot the dependence of the three detected precession frequencies on the external magnetic field applied along the [010] and [110] crystal axes. At saturation fields, which for the 7\% Mn-doped sample  are $\gtrsim70$~mT, the equilibrium magnetization vector is aligned with $H_{ext}$ and Eq.~(\ref{omega}) with $\varphi=\varphi_H$ can be used to fit the data. We emphasize that all six displayed dependences $f_n(H_{ext})$ for $n = 0$, 1, and 2,  and $\varphi_H = 45^\circ$ and 90$^\circ$ can be accurately fitted by one set of magnetic anisotropy constants. We can therefore use Eq.~(\ref{Hn}) to convert the measured frequency spacing of individual SWRs to $\Delta H_n$. In Fig.~5d we show that $\Delta H_n$ in our optimized epilayers is proportional to  $n^2$ as expected for the Kittel modes in homogeneous films. 

\begin{figure}[h!]
%\vspace*{-0.2cm}
\hspace{0cm}\includegraphics[width=.7\columnwidth,angle=0]{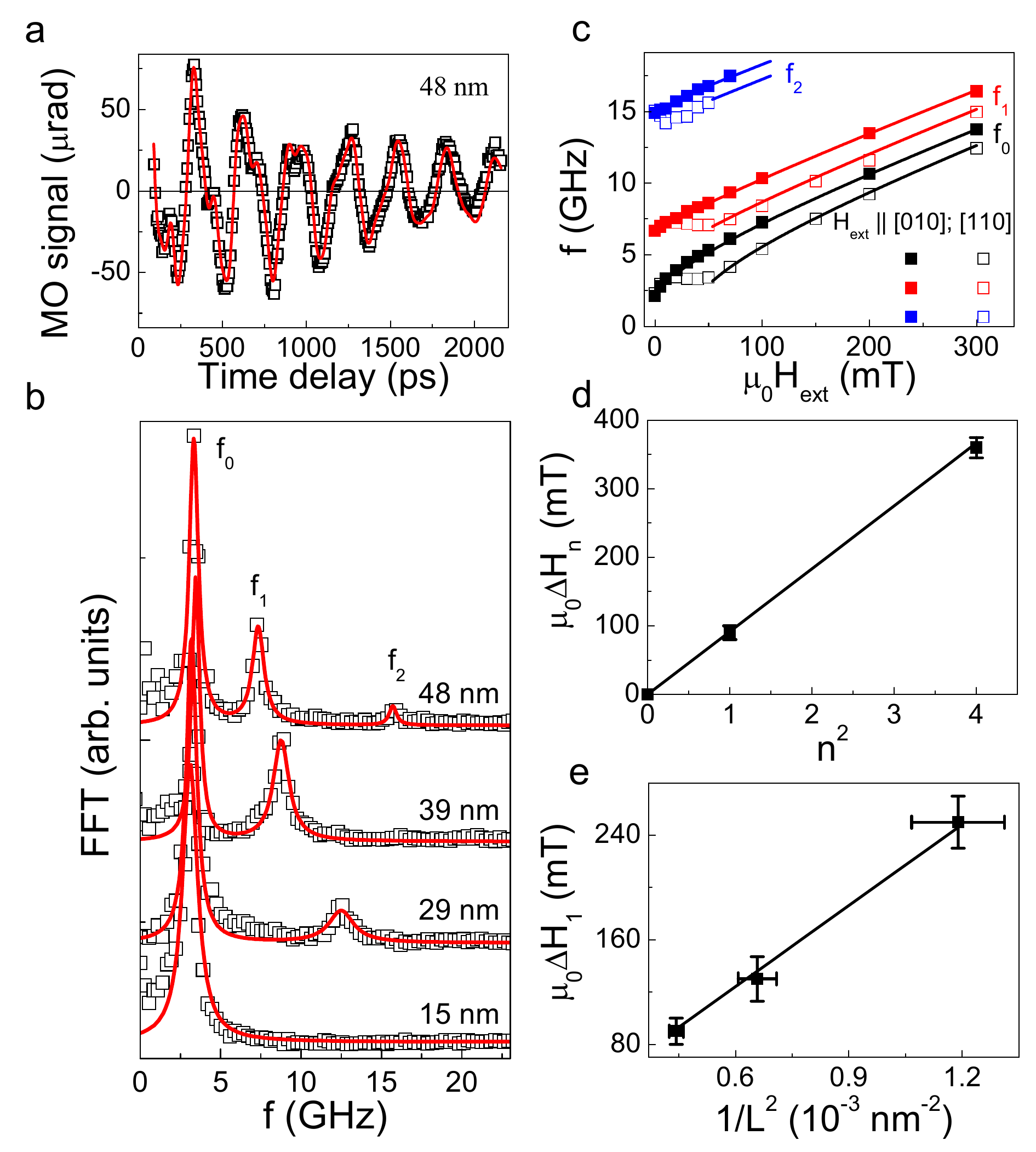}
%\vspace*{-0.5cm}
\caption{{\bf Determination of the spin stiffness constant from MO experiments.} {\bf a,} Oscillatory part of the MO signal  (points) measured in a 48~nm thick epilayer with 7\% nominal Mn doping for external magnetic field $\mu_0 H_{ext} = 20$~mT applied along the [010] crystallographic direction; line is a fit by a sum of three damped harmonic functions. {\bf b,} Fourier spectra of oscillatory MO signals  (points) measured for  $\mu_0 H_{ext} = 20$~mT  applied along the [010] crystallographic direction in samples prepared by etching from the 48~nm thick epilayer. The curves are labeled by the film thicknesses, normalized, and vertically off-set for clarity; lines are fits by a sum of Lorentzian peaks. {\bf c,} Dependence of the measured precession frequency (points) on the magnetic field for two different orientations of the field in the 48~nm thick epilayer; lines are fits by Eq.~(1). {\bf d,} Dependence of the measured mode spacing on square of the mode number in the 48~nm thick epilayer. {\bf e,} Dependence of the spacing between the two lowest modes ($\Delta H_1$) on the film thickness. Lines in {\bf d} and {\bf e} are fits by Eq.~(2) with spin stiffness $D = 2.43$~meV\,nm$^2$.}
\label{fig5}
\end{figure}

The magnetic homogeneity and the applicability of Eq.~(\ref{Hn}) in our epilayers is further confirmed by the following experiments: We prepared three samples by etching the original 48 nm thick (Ga,Mn)As film down to the thicknesses of 39, 29 and 15~nm, respectively. As seen in Fig.~5b, the frequency $f_0$ is independent of the film thickness which confirms that it corresponds to the uniform precession mode and that the film is homogeneous, i.e., the magnetic anisotropy constants do not vary across the width of the (Ga,Mn)As epilayer.
The spacing $\Delta H_1$ shown in Fig.~5e scales as $L^{-2}$ and the values of $D$ extracted from  the $n$-dependence of the resonant field spacings in the $L=48$~nm epilayer (see Fig. 5d) and from the $L$-dependence  of  $\Delta H_1$ (see Fig. 5e) give the same $D = 2.43\pm0.15$~meVnm$^2$. Identical value of the spin stiffness was  also obtained from measurements in an epilayer grown with the same doping and thickness of 18~nm in which we detected the frequencies $f_0$ and $f_1$ and applied Eq.~(\ref{Hn}). These measurements confirm the reliability of extracted values of the spin stiffness. We note that the SWR frequencies are determined with high accuracy in our measurements and that the indicated error bars in Fig.~1d reflect the uncertainty of the film thickness. As shown in Fig.~1d, we observe a consistent, weakly increasing trend in $D$ with increasing doping and values of $D$ between 2 and 3~meVnm$^2$ in the studied ferromagnetic samples with nominal doping 3.8-9\%. (Note that apart from the difficulty of exciting magnetization precession in the very low and high-doped samples with stable easy-axes, the measurements of $D$ were unfeasible on the lower doping side of the series because of the increasing damping and the corresponding inability to detect the higher SWR modes.) 
Similar to the Gilbert damping constant, our measured spin stiffness constant in the optimized (Ga,Mn)As epilayers is comparable to the spin stiffness in conventional transition metal ferromagnets.\cite{Collins:1969_a}   

We remark, that we tested the inapplicability of the SWR experiments for the direct determination of the spin stiffness in thick non-uniform materials. In the Supplementary information we show measurements in $\sim 500$~nm thick as-grown and annealed samples with 7\% nominal Mn-doping. The Curie temperatures of $\sim 60$ and 90~K  can be inferred only approximately from smeared out singularities in $d\rho/dT$ and $M(T)$ and  are significantly smaller than $T_c$ in the thin optimized epilayers with the same nominal doping. The films are therefore clearly inhomogeneous and contain compensating defects. Because of the large thickness of the epilayers we observe up to five SWR modes, however, consistent with the inhomogeneous structure of the films,  the corresponding $\Delta H_n$ do not show the quadratic scaling with $n$ of the Kittel modes of Eq.~(\ref{Hn}).  

In the experiments discussed above we have established the systematic semiconducting doping trends and basic magnetic characteristics of epilayers which have been optimized to represent as close as possible the intrinsic properties of idealized, uniform and uncompensated (Ga,Mn)As. Our study supports the overall view of (Ga,Mn)As as a well behaved and understood degenerate semiconductor and band ferromagnet and, therefore, an ideal model system for spintronics research.  We conclude in this paragraph by commenting on the implications of systematic studies of optimized (Ga,Mn)As materials in the context of  the recurring alternative proposal of an intricate impurity band nature of conduction and magnetism of (Ga,Mn)As.\cite{Samarth:2012_a} In the impurity band picture, the Fermi level in materials with $\sim 10^{21}$~cm$^{-3}$ Mn-acceptor densities is assumed to reside in a narrow impurity band detached from the valence band, i.e., the band structure keeps the form closely reminiscent of a single isolated Mn$_{\rm Ga}$ impurity level. Previously, the systematic measurements of the infrared conductivity on the extensive set of optimized materials\cite{Jungwirth:2010_b} disproved one of the founding elements of the impurity band picture which was the red-shift  of the mid-infrared peak with increasing doping.\cite{Burch:2006_a} In the systematic measurements in Ref.~\onlinecite{Jungwirth:2010_b}, the  mid-infrared peak was observed to blue-shift\cite{Jungwirth:2010_b,Chapler:2011_a}  and experimentalists focusing on the infrared spectroscopy\cite{Acbas:2009_a,Jungwirth:2010_b,Chapler:2011_a} reached the consensus that the valence and impurity bands are merged in the highly doped ferromagnetic (Ga,Mn)As materials.  The large values of the spin stiffness of the order meVnm$^2$, experimentally determined in the present work,  are consistent with model Hamiltonian and {\em ab initio} calculations\cite{Konig:2001_a,Brey:2003_a,Bouzerar:2006_c,Werpachowska:2010_a}  which all consider or obtain the band structure of the ferromagnetic (Ga,Mn)As with merged valence and impurity bands.\cite{Masek:2010_a} On the other hand, for carriers localized in a narrow impurity band the expected spin stiffness would be small in a dilute moment system like (Ga,Mn)As, in which the magnetic coupling between remote Mn moments is mediated by the carriers.\cite{MacDonald:2005_a} By recognizing that the bands are merged, the distinction between a "valence" and "impurity" band picture of ferromagnetic (Ga,Mn)As becomes mere semantics with no fundamental physics relevance. Simultaneously, it is important to keep in mind that the moderate acceptor binding energy of Mn$_{\rm Ga}$ shifts the insulator-to-metal transition to orders of magnitude higher doping densities than in the case of common shallow non-magnetic acceptors.\cite{Jungwirth:2007_a,Masek:2010_a} Disorder and correlation effects, therefore, play a comparatively more significant role in (Ga,Mn)As than in  degenerate semiconductors with common shallow dopants and any simplified one-particle band picture of ferromagnetic (Ga,Mn)As can only represent a proxy to the electronic structure of the material.

\section*{Acknowledgment}
We acknowledge theoretical assistence of Pavel Motloch and support  from EU ERC Advanced Grant No. 268066 and FP7-215368 SemiSpinNet, from the Ministry of Education of the Czech Republic Grants No. LM2011026, from the Grant Agency of the Czech Republic Grant No. 202/09/H041 and P204/12/0853, from the Charles University in Prague Grant No. SVV-2012-265306 and 443011, from the Academy of Sciences of the Czech Republic Preamium Academiae, and from U.S. grants ONR-N000141110780, NSF-MRSEC DMR-0820414, NSF-DMR-1105512.
.

\protect\newpage

\vspace*{1cm}


\section*{Supplementary material}

# Establishing micromagnetic parameters of ferromagnetic semiconductor (Ga,Mn)As: Supplementary information

P. Němec,[1] V. Novák,[2] N. Tesařová,[1] E. Rozkotová,[1] H. Reichlová[2,1], D. Butkovičová[1], F. Trojánek,[1] K. Olejník,[2] P. Malý,[1] R. P. Campion,[3] B. L. Gallagher,[3] Jairo Sinova,[4,2] and T. Jungwirth[2,3]

[1] *Faculty of Mathematics and Physics, Charles University in Prague, Ke Karlovu 3, 121 16 Prague 2, Czech Republic*

[2] *Institute of Physics ASCR, v.v.i., Cukrovarnická 10, 16253 Praha 6, Czech Republic*

[3] *School of Physics and Astronomy, University of Nottingham, Nottingham NG72RD, United Kingdom*

[4] *Department of Physics, Texas A&M University, College Station, TX 77843-4242, USA*

## EXPERIMENTS PRECEEDING OUR DETERMINATION OF MICROMAGNETIC PARAMETERS OF (GA,MN)AS

Magnetic anisotropy fields, Gilbert damping constant and spin-stiffness are the basic parameters of a ferromagnet which determine its micromagnetic properties. The anisotropy fields are associated with the energy required to coherently rotate magnetic moments of the entire ferromagnet. They can be determined in a straightforward way in magnetization or magneto-transport measurements from external magnetic fields required to reorient the magnetization of a ferromagnetic sample, or in magnetization dynamics experiments from the field-dependent resonant frequencies [1, 2, 3, 4].

Gilbert damping characterizes dissipative processes that drive the magnetization motion towards an equilibrium state. This phenomenon is usually investigated by the frequency-domain-based ferromagnetic resonance (FMR) experiment where the phenomenological Gilbert damping coefficient $\alpha$ is deduced from the resonance peak linewidth [3, 4]. The experimentally measured FMR linewidths contain not only the frequency-dependent linewidth due to the Gilbert damping but also the frequency-independent inhomogeneous linewidth broadening [3, 4]. To separate them, it is necessary to measure the linewidths at several microwave frequencies [3, 4]. In FMR these frequencies are given by the resonant-cavity frequency that significantly complicates the frequency change.

Therefore, the experiments are usually performed at only two different frequencies (typically, 9 and 35 GHz [4]) that makes the corresponding separation of the individual components in the measured signal rather questionable. Alternatively, $\alpha$ can be determined from the time-domain based magneto-optical pump-and-probe experiment by fitting the damping of the measured oscillatory data by Landau-Lifshitz-Gilbert equation [5, 6]. However, to obtain the Gilbert damping coefficient from the measured value of $\alpha$ it is necessary to take into account a realistic magnetic anisotropy of the investigated samples (see below). Moreover, also the frequency dependence of $\alpha$ has to be measured for a separation of the intrinsic value of the Gilbert damping coefficient from the inhomogeneous parts of $\alpha$. The absence of these two requirements and the un-optimized magnetic properties of the investigated samples led to a large scatter in the deduced values of $\alpha$ for $Ga_{1-x}Mn_xAs$ with a different Mn content $x$: The increase of $\alpha$ from $\approx 0.02$ to $\approx 0.08$ for the increase of $x$ from 3.6% to 7.5% was reported in Ref. 5. On the contrary, in Ref. 6 the values of $\alpha$ from 0.06 to 0.19 – without any apparent doping trend – were observed for $x$ from 2% to 11%.

The spin-stiffness is associated with the exchange energy of non-uniform local directions of the magnetization, in particular with the energy of small wave-vector spin-wave excitations of the ferromagnet. Considering a specific model of thermodynamic properties of the studied ferromagnet, the spin-stiffness can be indirectly inferred from the measured temperature dependence of magnetization [7], Curie temperature [7], or domain wall width [8]. The direct determination of the spin-stiffness from magnetization dynamics experiments is significantly more challenging than in the case of the magnetic anisotropy fields [9-13]. The low-energy non-uniform collective excitations of the system can be strongly affected by inhomogeneities or surface properties of the ferromagnet for which specific models have to be assumed in order to extract the spin-stiffness constant from the measured data. An exception are the Kittel spin-wave modes of a uniform thin-film ferromagnet for which the spin-stiffness parameter $D$ is directly obtained from the measured resonant fields (see below). To date, spin-wave resonance measurements of (Ga,Mn)As have been reported on > 100 nm thick epilayers [9-12]. The Kittel modes with $\Delta H_n \sim n^2$ were observed only in a 120 nm thick, 8% Mn doped (Ga,Mn)As for magnetic fields applied close to the magnetic easy-axis [11]. Measurements of the same sample in other field orientations showed different trends which indicated the presence of strong inhomogeneities and surface dependent effects [11]. A non-Kittel-like linear or sublinear dependence of the resonant fields on the mode index has been reported also in the other ferromagnetic resonance measurements of thick (Ga,Mn)As epilayers [9-12]. In

complementary studies of the magnetization dynamics induced and detected by magneto-optical pump-and-probe measurements, only two resonant frequencies were identified [13]. Based on the theoretical modeling, they were not ascribed to the Kittel modes but rather to coupled bulk-surface modes which again made the extraction of the spin-stiffness constant dependent on the considered model of bulk and surface properties of the studied sample [13]. The extracted values of the spin-stiffness from all available magnetic resonance data in (Ga,Mn)As materials, complemented by values inferred from magnetization and domain studies [7, 8], are scattered over more than an order of magnitude and show no clear trend as a function of Mn-doping or other material parameters of the (Ga,Mn)As ferromagnetic semiconductor [14].

In this Supplementary material we show how we are able to deduce from a *single* magneto-optical pump-and-probe experiment all these micromagnetic parameters. In particular, the anisotropy fields can be determined from the dependence of the precession frequency on the external magnetic filed combined with the probe-polarization dependence of the precession signal amplitude. The Gilbert damping constant can be deduced from the precession signal damping. Finally, the spin stiffness can be obtained from the mutual spacing of the precession modes which are present in the measured oscillatory magneto-optical signal.

## SAMPLES

The time-resolved magneto-optical experiments described below were performed in a large set of optimized (Ga,Mn)As epilayers whose selected properties are described in detail in the main paper. In Fig. 1 we show results of the Hall effect measurements at 4.2 K. For this purpose the samples were lithographically patterned into Hall-bars of 60 μm width. It can be seen in the figure that the Hall signal is affected by longitudinal magnetoresistance $R_{xx}$ of the samples, especially at low dopings. Therefore, we extracted $p$ from high field data and by fitting the measured transversal resistance $R_{xy}$ by

$$R_{xy} = B/(epd) + k_1 R_{xx} + k_2 R_{xx}^2 \qquad (1)$$

where $d$ is the sample thickness and $k_1$ and $k_2$ are fitting constants reflecting the anomalous Hall effect and possible imperfections in the geometry of the Hall bars. We also emphasize that, apart from the common experimental scatter and from the corrections due to the non-zero magnetoresistance and due to the anomalous Hall effect, the carrier density can in principle be

inferred only approximately from the slope of the Hall curve in a multi-band, spin-orbit coupled exchange-split system such as the (Ga,Mn)As. The error bar due to the multi-band nature is estimated to be ≈20% [15]. Due to these uncertainties we can only make semi-quantitative conclusions based on the measured Hall effect hole densities.

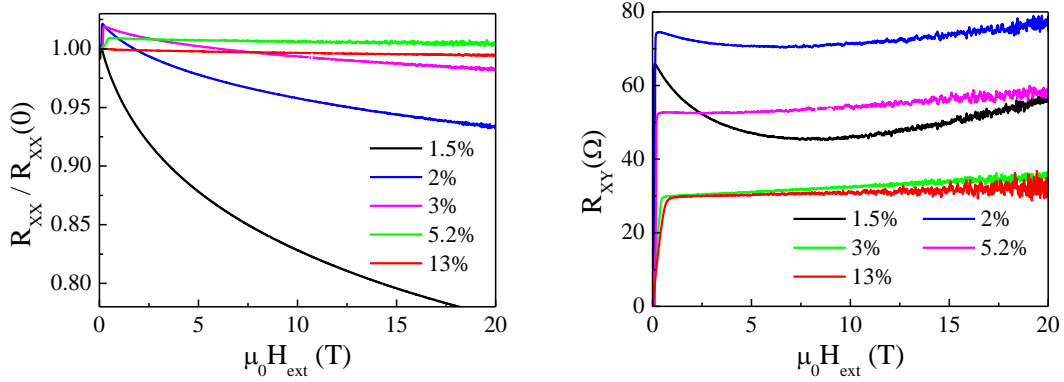

Fig. 1. (a) Longitudinal resistances $R_{xx}$ [normalized to $R_{xx}(0)$], and (b) transversal (Hall) resistances $R_{xy}$ as a function of normal magnetic field $\mu_0 H_{ext}$ measured in (Ga,Mn)As epilayer with depicted Mn concentration $x$; samples temperature 4.2 K.

For an evaluation of material parameters from an experimentally measured data (e.g., for an evaluation of the hole densities from the measured transversal resistances which is described above) it is necessary to know the (Ga,Mn)As epilayer thicknesses. However, accurate determination of layer thicknesses is a nontrivial task in case of thin (Ga,Mn)As layers. Some standard techniques (e.g., X-ray reflectivity or optical ellipsometry) are inapplicable due to the weak contrast between the (Ga,Mn)As layer and the GaAs substrate, or unknown optical parameters. The relative accuracy of other common techniques (e.g., of X-ray diffraction) does not exceed 10% because of the small thickness of the measured layer. Therefore, we used a thickness estimation based on the following quantities: (i) the growth time and the growth rate of the GaAs buffer layer measured by the RHEED oscillations (typical accuracy of ±3%); (ii) increase in the growth rate by adding the known Mn-flux measured by the beam-flux monitor relatively to the Ga flux (typical accuracy of ±5% of the Mn vs. Ga flux ratio); (iii) reduction of thickness by the native oxidation (-1.5 nm ± 0.5 nm); (iv) reduction of thickness by thermal oxidation (-1.0 nm ± 0.5 nm). Relative accuracy of steps (i) and (ii) was verified on separate calibration growths of (Ga,Mn)As on AlAs, where an accurate X-ray reflectivity method to measure the (Ga,Mn)As layer thickness could be used. Typical thicknesses of the native and the thermal oxides in steps (iii) and (iv) were

determined by XPS. The resulting total accuracy of the (Ga,Mn)As layer thickness determination is thus 3% (relative random error) and 1 nm (systematic error).

## EXPERIMENTAL DETAILS ABOUT MAGNETO-OPTICAL EXPERIMENTS

We investigated laser-pulse induced dynamics of magnetization by a pump-and-probe magneto-optical (MO) technique. A schematic diagram of the experimental set-up is shown in Fig. 2. The output of a femtosecond laser is divided into a strong pump pulse and a weak probe pulse that are focused to a same spot on the sample. Laser pulses, with the time width of 200 fs and the repetition rate of 82 MHz, were tuned to 1.64 eV, i.e. above the semiconductor band gap, in order to excite magnetization dynamics by photon absorption. The pump pulses were usually circularly polarized (with a helicity controlled by a wave plate) and the probe pulses were linearly polarized. The measured magneto-optical signals correspond to the probe polarization rotation induced by the pump pulses (see Fig. 2). The experiment was performed close to the normal incidence geometry ($\theta_i = 2°$ and $8°$ for pump and probe pulses, respectively) with a sample mounted in a cryostat, which was placed between the poles of an electromagnet. All the experimental data in this Supplementary material were measured at temperature of 15 K, at pump excitation intensity 30 -40 µJ.cm$^{-2}$, and they correspond to the helicity-independent part of the measured signal [16]. The external magnetic field $H_{ext}$ was applied in the sample plane at an angle $\varphi_H$ with respect to the [100] crystallographic direction in the sample plane (see Fig. 2). Prior to all time-resolved experiments, we always prepared the magnetization in a well-defined state by first applying a strong saturating magnetic field at an angle $\varphi_H$ and then reducing it to the desired magnitude of $H_{ext}$.

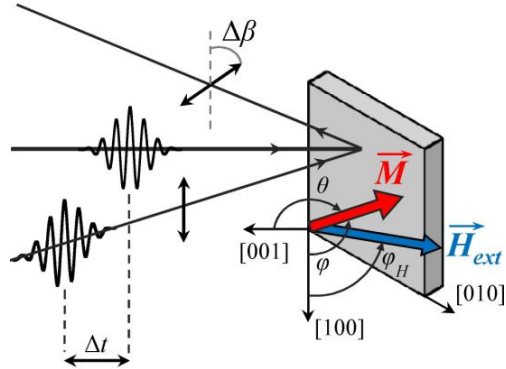

Fig. 2. Schematic diagram of the experimental set-up for a detection of the magnetization precession induced by an impact of the circularly polarized femtosecond laser pump pulse in (Ga,Mn)As. Rotation of the polarization plane of reflected linearly polarized probe pulses is measured as a function of the time delay $\Delta t$ between pump and probe pulses. The orientation of magnetization in the sample is described by the polar angle $\varphi$ and azimuthal angle $\theta$. The external magnetic field $H_{ext}$ is applied in the sample plane at an angle $\varphi_H$.

There are several microscopic mechanisms that can lead to a precession of magnetization due to the impact of pump laser pulse. In particular, very recently we reported on the precession of magnetization due to optical spin-transfer torque (OSTT) [16] and optical spin-orbital torque (OSOT) [17]. However, the most common mechanism, which is responsible for the oscillatory MO signals measured in the majority of (Ga,Mn)As samples at low excitation intensities, is the change of the sample magnetic anisotropy due to the pump-induced temperature increase [17] that is schematically shown in Fig. 3. Before an impact of the pump pulse the magnetization points to the easy axis direction [see Fig. 3(a)]. Absorption of the laser pulse leads to a photo-injection of electron-hole pairs. The subsequent fast nonradiative recombination of photo-injected electrons induces a transient increase of the lattice temperature (within tens of picoseconds after the impact of the pump pulse). The laser-induced change of the lattice temperature then leads to a change of the easy axis position [17]. As a result, magnetization starts to follow the easy axis shift by the precessional motion [see Fig. 3(b)]. Finally, dissipation of the heat and recombination of the excess holes lead to the return of the easy axis to the equilibrium position and the precession of magnetization is stopped by the Gilbert damping [see Fig. 3(c)]. The most important point from the perspective of the present paper is that the precession of magnetization induced by the laser pulses is determined by the magnetic anisotropy of the sample which makes this method an all-optical analog to FMR [18].

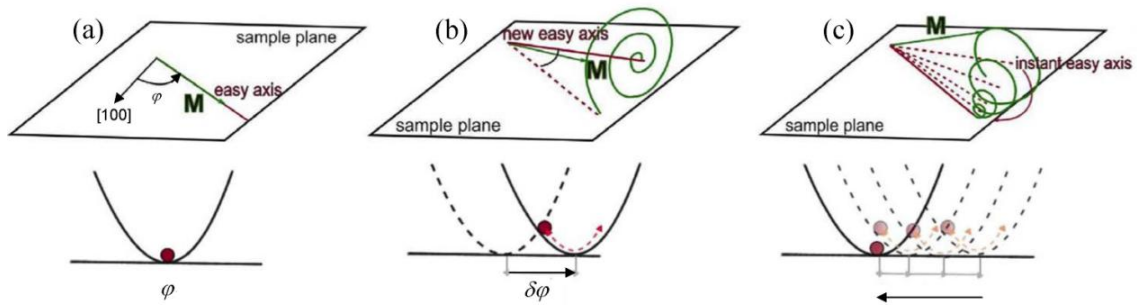

Fig. 3. Schematic illustration of the thermal laser pulse-induced precession of magnetization. (a) In the equilibrium, the magnetization points to the easy axis direction, which is located in the sample plane at azimuthal angle $\varphi$. (b) Impact of a pump pulse induces a transient increase of the lattice temperature that leads to a change of the easy axis position and, consequently, to the precession of magnetization. (c) Dissipation of the heat leads to the return of the easy axis to the equilibrium position. Simultaneously with this, the precession of magnetization is stopped by the Gilbert damping.

## ANALYTICAL DESCRIPTION OF MAGNETIZATION DYNAMICS IN (GA,MN)AS

The dynamics of magnetization is described by the Landau-Lifshitz-Gilbert (LLG) equation. We used LLG equation in spherical coordinates where the time evolution of magnetization magnitude $M_s$ and orientation, which is characterized by the polar $\theta$ and azimuthal $\varphi$ angles, is given by:

$$\frac{dM_s}{dt} = 0, \tag{2}$$

$$\frac{d\theta}{dt} = -\frac{\gamma}{(1+\alpha^2)M_s}\left(\alpha \cdot A + \frac{B}{\sin\theta}\right), \tag{3}$$

$$\frac{d\varphi}{dt} = \frac{\gamma}{(1+\alpha^2)M_s \sin\theta}\left(A - \frac{\alpha \cdot B}{\sin\theta}\right), \tag{4}$$

where $\alpha$ is the Gilbert damping coefficient. The gyromagnetic ratio $\gamma = (g\mu_B)/\hbar$, where $g = 2$ is the Landé g-factor of Mn moments, $\mu_B$ the Bohr magneton, and $\hbar$ is the reduced Planck constant. Functions $A = dF/d\theta$ and $B = dF/d\varphi$ are the derivatives of the energy density functional $F$ with respect to $\theta$ and $\varphi$, respectively. We expressed $F$ in a form [2]:

$$F = M\left[\begin{array}{l} K_c \sin^2\theta\left(\frac{1}{4}\sin^2 2\varphi \sin^2\theta + \cos^2\theta\right) - K_{out}\cos^2\theta - \frac{K_u}{2}\sin^2\theta(1-\sin 2\varphi) - \\ -H_{ext}[\cos\theta\cos\theta_H + \sin\theta\sin\theta_H \cos(\varphi-\varphi_H)] \end{array}\right], \tag{5}$$

where $K_c$, $K_u$ and $K_{out}$ are constants that characterize the cubic, uniaxial and out-of-plane magnetic anisotropy fields in (Ga,Mn)As, respectively, and $H_{ext}$ is the external magnetic field whose orientation is given by the angles $\theta_H$ and $\varphi_H$. For a small deviations $\delta\theta$ and $\delta\varphi$ from the equilibrium values $\theta_0$ and $\varphi_0$, the solution of Eqs. (3) and (4) can be written in a form

$$\theta(t) = \theta_0 + A_\theta e^{-k_D t}\cos(2\pi f t + \Phi_\theta), \tag{6}$$

$$\varphi(t) = \varphi_0 + A_\varphi e^{-k_D t}\cos(2\pi f t + \Phi_\varphi), \tag{7}$$

where the constants $A_\theta$ ($A_\varphi$) and $\Phi_\theta$ ($\Phi_\varphi$) describe the initial amplitude and phase of $\theta$ ($\varphi$), respectively. The precession frequency $f$ and oscillation damping rate $k_D$ are given by

$$f = \frac{g\mu_B}{h(1+\alpha^2)} \sqrt{\begin{array}{l}\left(H_{ext}\cos(\varphi-\varphi_H)-2K_{out}+\frac{K_c(3+\cos 4\varphi)}{2}+2K_u\sin^2\left(\varphi-\frac{\pi}{4}\right)\right)\times \\ \times\left(H_{ext}\cos(\varphi-\varphi_H)+2K_c\cos 4\varphi-2K_u\sin(2\varphi)\right)+ \\ +\alpha^2\left\{\begin{array}{l}\left(H_{ext}\cos(\varphi-\varphi_H)-2K_{out}+\frac{K_c}{2}(3+\cos 4\varphi)+2K_u\sin^2\left(\varphi-\frac{\pi}{4}\right)\right)\times \\ \times\left(H_{ext}\cos(\varphi-\varphi_H)+2K_c\cos 4\varphi-2K_u\sin(2\varphi)\right)- \\ \left(H_{ext}\cos(\varphi-\varphi_H)-K_{out}+2K_c\frac{1}{8}(3+5\cos 4\varphi)+\frac{K_u}{2}(1-3\sin(2\varphi))\right)^2\end{array}\right\}\end{array}},$$

(8)

$$k_D = \alpha \frac{g\mu_B}{2\hbar(1+\alpha^2)}\left(2H_{ext}\cos(\varphi-\varphi_H)-2K_{out}+\frac{K_c}{2}(3+5\cos 4\varphi)+K_u(1-3\sin(2\varphi))\right). \quad (9)$$

In our case, the investigated (Ga,Mn)As epilayers are in-plane magnets (i.e., $\theta \approx \pi/2$), the external magnetic field is applied in the sample plane (i.e., $\theta_H \approx \pi/2$), and the precession damping is relatively slow (i.e., $\alpha^2 \approx 0$) which yields

$$f = \frac{g\mu_B}{h}\sqrt{\begin{array}{l}\left(H_{ext}\cos(\varphi-\varphi_H)-2K_{out}+\frac{K_c(3+\cos 4\varphi)}{2}+2K_u\sin^2\left(\varphi-\frac{\pi}{4}\right)\right)\times \\ \times\left(H_{ext}\cos(\varphi-\varphi_H)+2K_c\cos 4\varphi-2K_u\sin(2\varphi)\right)\end{array}}, \quad (10)$$

$$k_D = \alpha \frac{g\mu_B}{2\hbar}\left(2H_{ext}\cos(\varphi-\varphi_H)-2K_{out}+\frac{K_c}{2}(3+5\cos 4\varphi)+K_u(1-3\sin(2\varphi))\right). \quad (11)$$

Eq. (10) express the sensitivity of the magnetization precession frequency to the magnetic anisotropy of the material that is a well-known effect which form the basis for the interpretation of FMR [3]. More interestingly, Eq. (11) shows that the precession damping $k_D$, which is measured experimentally, depends not only on the Gilbert damping parametr $\alpha$ but also on the sample anisotropy and on the mutual orientation of the external magnetic field and the magnetization.

We note that in previously reported magneto-optical pump-and-probe experiments [5, 6, 19-23] the measured experimental data were modeled by LLG equation in the form

$$\frac{d\vec{M}(t)}{dt} = -\gamma \left[ \vec{M}(t) \times \vec{H}_{eff}(t) \right] + \frac{\alpha}{M_s} \left[ \vec{M}(t) \times \frac{d\vec{M}(t)}{dt} \right], \tag{12}$$

where $\vec{H}_{eff}$ is the effective magnetic field. However, in (Ga,Mn)As the magnetic anisotropy is rather complex and, therefore, modeling of MO signals by LLG in this form does not provide realistic values of $\alpha$ because it is not possible to disentangle the effect of magnetic anisotropy from $\alpha$ [see Eq. (11)]. We believe that this is one of the reasons why the dependence of $\alpha$ on Mn concentration was so different in Ref. 5 and Ref 6. Similarly, the change of magnetic anisotropy of (Ga,Mn)As during the deposition of metal overlayer could be partially responsible for the changes of $\alpha$ that were reported in Ref. 23.

## EVALUATION OF MAGNETIC ANISOTROPY

The dependence of the precession frequency on the magnetic anisotropy fields and on the magnitude and orientation of external magnetic field [cf. Eq. (10)] enables to evaluate the magnetic anisotropy from the experimentally measured precession frequencies very similarly as in the case of FMR [3]. In particular, for a sufficiently strong external magnetic field $\varphi = \varphi_H$ and the following equations can be used to fit the precession frequencies measured

a) for $H_{ext}$ along the [110] crystallographic direction (i.e., $\varphi_H = \pi/4$):

$$f = \frac{g\mu_B}{h} \sqrt{(H_{ext} - 2K_{out} + K_c)(H_{ext} - 2K_c - 2K_u)} \tag{13}$$

b) for $H_{ext}$ along the [010] crystallographic direction (i.e., $\varphi_H = \pi/2$):

$$f = \frac{g\mu_B}{h} \sqrt{(H_{ext} - 2K_{out} + 2K_c + K_u)(H_{ext} + 2K_c)} \tag{14}$$

c) for $H_{ext}$ along the [-110] crystallographic direction (i.e., $\varphi_H = 3\pi/4$):

$$f = \frac{g\mu_B}{h} \sqrt{(H_{ext} - 2K_{out} + K_c + 2K_u)(H_{ext} - 2K_c + 2K_u)} \tag{15}$$

As an example, in Fig. 4 we show the measured dependences $f(H_{ext})$ and their fits for two orientations of $H_{ext}$. To increase the precision of the magnetic anisotropy determination even further, for all the investigated samples we supplemented this method by two additional

experimental techniques that provide information about the samples magnetic anisotropy – namely, the probe-polarization dependence of the MO precession signal amplitude and SQUID magnetometry.

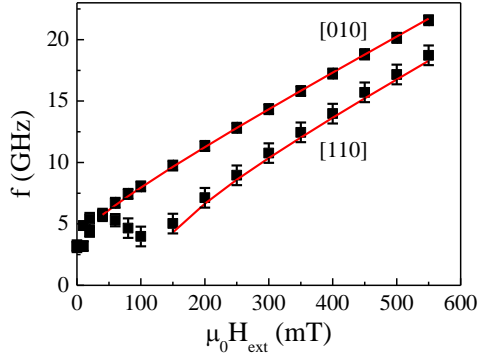

Fig. 4. Dependence of the precession frequency $f$ on external magnetic field $H_{ext}$ applied along the [010] and [110] crystallographic directions in $Ga_{1-x}Mn_xAs$ epilayer with $x$ = 5.2% (points); the lines are fits by Eqs. (14) and (13), respectively, with $K_c$ = 31 mT, $K_u$ = 27.5 mT, and $K_{out}$ = -190 mT.

In (Ga,Mn)As there two MO effects that are responsible for the measured rotation of the polarization plane $\Delta\beta$ of the reflected linearly polarized light at normal incidence [24]. The first of the MO effects is the well-known polar Kerr effect (PKE), where $\Delta\beta$ occurs due to the different index of refraction for $\sigma^+$ and $\sigma^-$ circularly polarized light propagating parallel to the direction of magnetization $M$. The polarization rotation due to PKE is proportional to the projection of magnetization to the direction of light propagation, it is linear in magnetization (i.e., its sign is changed when the direction of magnetization is reversed), and it is independent on the orientation of the input linear polarization $\beta$ (see Fig. 5(b) for the angle definition) [24]. The second MO effect is the magnetic linear dichroism (MLD), which originates from different absorption (reflection) coefficient for light linearly polarized parallel and perpendicular to $M$, that occurs if the light propagates perpendicular to the direction of magnetization $M$. The polarization rotation due to MLD is proportional to the projection of magnetization to the direction perpendicular to the direction of light propagation, it is quadratic in magnetization (i.e., its sign is not changed when the direction of magnetization is reversed) and it varies as $\sin(2\beta)$ [24]. In Fig. 5(a) we show the MO signals measured by probe pulses with different orientations $\beta$ for identical pumping conditions. The measured dynamical MO signal $\delta MO$, which is a function of the time delay between pump and probe pulses $\Delta t$ and the probe polarization orientation $\beta$, can be fitted well by the phenomenological equation [24],

$$\delta MO(\Delta t, \beta) = A(\beta)\cos[2\pi f \Delta t + \Phi(\beta)]e^{-\Delta t/\tau_G} + C(\beta)e^{-\Delta t/\tau_p}, \qquad (16)$$

where *A* and *C* are the amplitudes of the oscillatory and pulse function, respectively, *f* is the ferromagnetic moment precession frequency, $\Phi$ is the phase factor, $\tau_G$ is the Gilbert damping time, and $\tau_p$ is the pulse function decay time. All the measured data in Fig. 5(a) can be fitted well by Eq. (16) with a one set of parameters *f*, $\tau_G$ and $\tau_p$. The dependence $A(\beta)$ obtained by

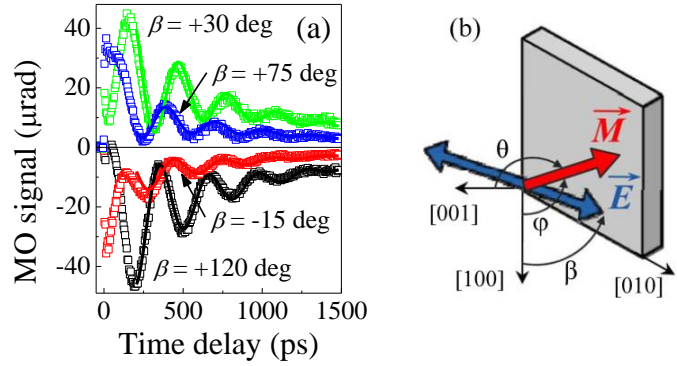

Fig. 5. (a) Dynamics of the MO signal measured by probe pulses with different probe polarization orientations $\beta$ in (Ga,Mn)As epilayer with *x* = 5.2% for $\mu_0 H_{ext}$ = 0 mT (points); lines are fits by Eq. (16) with parameters *f* = 3.2 GHz, $\tau_G$ = 360 ps and $\tau_p$ = 1050 ps. (b) Definition of the angle $\beta$ that describes the orientation of the probe polarization plane ***E***.

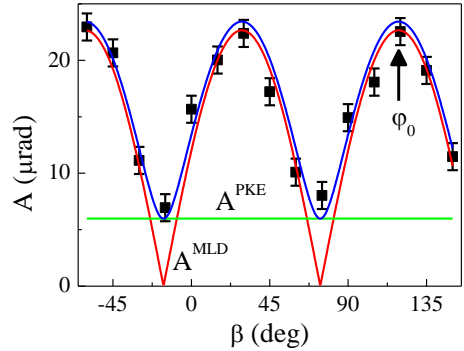

Fig. 6. Probe-polarization dependence of the oscillatory part *A* of the MO signal that was obtained by fitting the dynamics shown in Fig. 5(a) by Eq. (16); the values of *A* at time delay of 200 ps are shown (points). Lines are fits of $A(\beta)$ by a sum of a polarization-independent signal due to PKE and a polarization-dependent signal due to MLD (Eq. (2) in Ref. 24). The vertical arrow depicts the deduced easy axis position in the sample without the pump pulse, $\varphi_0$.

this fitting procedure is displayed in Fig. 6. The position of the maximum in the dependence $A(\beta)$ at $\beta \approx 120°$ corresponds to the equilibrium position of the easy axis in the sample - i.e., the in-plane position of magnetization without the pump pulse $\varphi_0$ [24]. The position of the easy axis in the sample plane is given by the relative magnitude of the cubic ($K_c$) and uniaxial

($K_u$) anisotropy fields. Therefore, by measuring $\varphi_0$ without external magnetic field applied, we are directly measuring the ratio $K_c / K_u$.

The in-plane anisotropy constants can be obtained also from magnetization loops measured by SQUID magnetometry. For any external magnetic field the orientation of magnetization is determined by the minimum of the energy [cf. Eq. (5)]. If the orientation of magnetization as function of external magnetic field is known, the projection of the magnetization into the measurement axis can be easily numerically evaluated for every point of the magnetization loop. To obtain the anisotropy constants, we fitted the experimental data measured by SQUID until we obtained the best agreement between the data and the calculated magnetization – see Fig. 7. It is worth noting that this model does not describe the switching mechanism (governed by the domain wall physics which is not treated in our single domain description), so the parts of hysteresis loops containing the switching were not used in the analysis. Moreover, in the case of uniaxial systems ($K_u > K_c$) an analytical expression for the magnetization measured along the hard axis can be utilized to analyze the data [1].

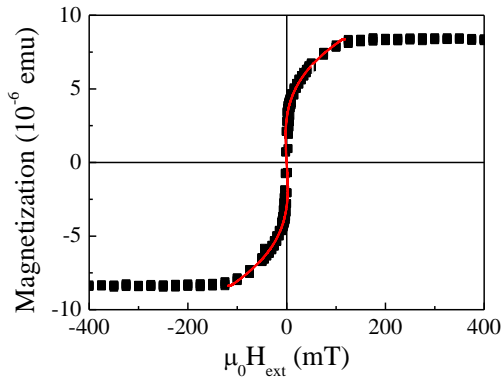

Fig. 7. Evaluation of the magnetic anisotropy from the SQUID magnetometry. The SQUID measurement along [110] crystallographic direction in (Ga,Mn)As epilayer with $x = 5.2\%$ (points) is compared with the calculated magnetization projection for $K_c = 31$ mT and $K_u = 27.5$ mT (line).

To sum up, by a simultaneous fitting of the measured dependence of the precession frequency on an external magnetic field (Fig. 4), of the MO signal precession amplitude on a probe-polarization (Fig. 6), and of the data measured by SQUID magnetometry (Fig. 7) we evaluated very precisely the magnetic anisotropy for all the investigated samples. The example of the obtained in-plane angular dependence of the free energy in (Ga,Mn)As epilayer with Mn concentration $x = 5.2\%$ is shown in Fig. 8.

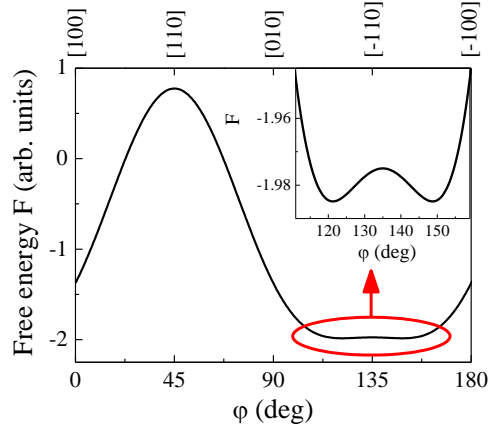

Fig. 8. In-plane angular dependence of the free energy [Eq. (5)] in (Ga,Mn)As epilayer with $x = 5.2\%$; anisotropy fields $K_c = 31$ mT and $K_u = 27.5$ mT.

## DETERMINATION OF GILBERT DAMPING COEFFICIENT

For numerical modeling of the measured MO data, we first computed from the LLG equation (Eqs. (3) and (4) with the measured magnetic anisotropy fileds) the time-dependent deviations of the spherical angles [$\delta\theta(t)$ and $\delta\varphi(t)$] from the corresponding equilibrium values ($\theta_0$, $\varphi_0$). Then we calculated how such changes of $\theta$ and $\varphi$ modify the static magneto-optical response of the sample $MO^{stat}$, which is the signal that we detect experimentally [24]:

$$\delta MO(\Delta t, \beta) = -\delta\theta(\Delta t)P^{PKE} + \delta\varphi(\Delta t)P^{MLD}2\cos 2(\varphi_0 - \beta) + \frac{\delta M_s(\Delta t)}{M_0}P^{MLD}2\sin 2(\varphi_0 - \beta). \quad (17)$$

The first two terms in Eq. (17) are connected with the out-of-plane and in-plane movement of magnetization, and the last term describes a change of the static magneto-optical response of the sample due to the laser-pulses induced demagnetization [24]. $P^{PKE}$ and $P^{MLD}$ are MO coefficients that describe the MO response of the sample which we measured independently in a static MO experiment for all the samples – see Fig. 9 for MO spectra measured in sample with $x = 5.2\%$.

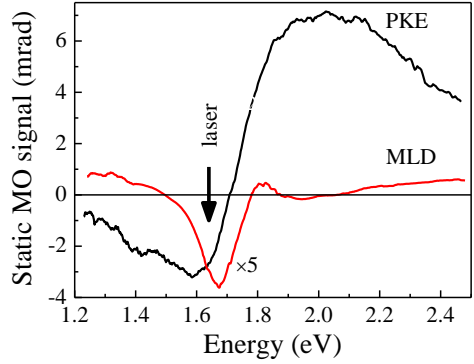

Fig. 9. Spectral dependence of static PKE and MLD in (Ga,Mn)As epilayer with $x = 5.2\%$, the arrow indicate the spectral position of the laser pulses used in the time-resolved experiment shown in Fig. 5 and Fig. 10; note that the data for MLD are multiplied by 5 for clarity.

The examples of the fitting of the dynamical MO optical data are shown in Fig. 10. The measured data can be fitted well by LLG for time delays longer than ≈ 150 ps, which is a time that it takes to establish the quasi-equilibrium conditions in the sample. We stress that the only fitting parameters in our modeling are the Gilbert damping coefficient $\alpha$, the initial deviation of the spherical angles from the corresponding equilibrium values, and the parameters describing the in-plane movement of the easy axis and the demagnetization signal, which are apparent as the non-oscillatory signal in the measured dynamics [24]. The obtained dependence of $\alpha$ on $H_{ext}$ is shown in Fig. 11(a) for two different orientations of $H_{ext}$. For $H_{ext}$ applied along the [010] direction, $\alpha$ decreases monotonously with $H_{ext}$. On the contrary, for $H_{ext}$ applied along [110] direction, $\alpha$ is a non-monotonous function of $H_{ext}$ reaching a similar values of $\alpha$ for 0 mT and 100 mT. However, this non-monotonous dependence is a consequence of the field-induced frequency decrease (see Fig. 4) when the magnetic field is applied along the magnetically hard [110] direction (see Fig. 8). When $\alpha$ is plotted as a function of the precession frequency (rather than the external field) we do not observe any significant difference between the different crystallographic directions – see Fig. 11(b). A field dependent damping parameter was reported in various magnetic materials and a variety of underlying mechanisms responsible for it were suggested as an explanation [25-29]. We note that the damping parameter $\alpha$ extracted from the fits should be regarded as a phenomenological parameter that accounts for combined effects of a (frequency independent) intrinsic Gilbert damping, an inhomogeneous broadening, a two magnon scattering, and various propagation spin wave processes resulting from the nonuniform spatial profile of the excited precession. We also note that the rate of decrease of $\alpha$ with $f$ is sample dependent and,

14

therefore, we show in the main paper the doping dependence of the frequency-independent part of $\alpha$.

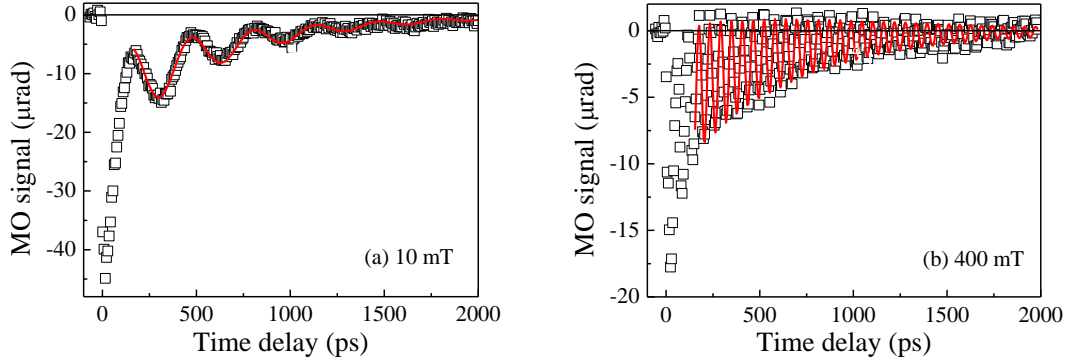

Fig. 10. Dynamics of the MO signal measured for external magnetic field (a) $\mu_0 H_{ext} = 10$ mT and (b) $\mu_0 H_{ext} = 400$ mT applied along the [010] crystallographic direction in (Ga,Mn)As epilayer $x = 5.2\%$ (points); lines are fits by LLG.

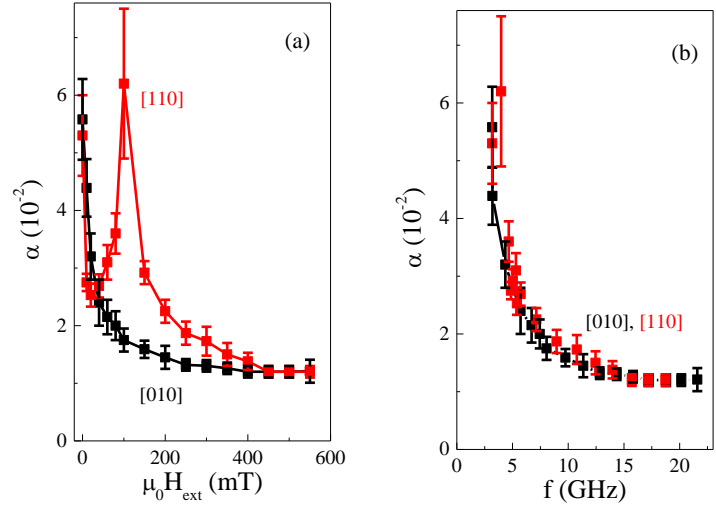

Fig. 11. (a) Dependence of the Gilbert damping coefficient $\alpha$ on external magnetic field $H_{ext}$ applied along the [010] and [110] crystallographic directions in (Ga,Mn)As epilayer with $x = 5.2\%$. (b) Same data as in (a) but as a function of the precession frequency $f$.

## DETERMINATION OF SPIN STIFFNESS

As we show in Fig. 5 of the main paper, we observed more than one precession mode in (Ga,Mn)As epilayers with a sufficient thickness. These precession modes are the spin wave resonances (SWRs) – i.e., spin waves (or magnons) that are selectively amplified by fulfilling the boundary conditions of the thin magnetic film [18, 30]. Up to now, SWRs in (Ga,Mn)As were investigated mainly in a frequency-domain where they are apparent as multiple absorption peaks in the FMR spectra [3, 10 - 12]. The existence of multiple

15

resonances in FMR reveal that there exist several external magnetic fields at which the Larmor precession frequency in the sample coincides with the microwave frequency. The resonant field for the $n$-th mode ($H_n$) is obtained by solving the LLG equation with a term corresponding to exchange interactions in the material and by considering the appropriate boundary condition [11]. In homogeneous thin films with a thickness $L$, only the perpendicular standing waves with a wave vector $k$ fulfilling the resonant condition $kL = n\pi$ are amplified; the mode with $n = 0$ denotes the uniform magnetization precession with zero $k$ vector. In principle, there exist two symmetric boundary conditions which are schematically illustrated in Fig. 12. The position of $n$-th SWR mode in the FMR spectrum $H_n$ is given by the Kittel relation [11] and the following equation applies

$$\Delta H_n \equiv H_0 - H_n = n^2 \frac{D}{g\mu_B} \frac{\pi^2}{L^2}, \tag{18}$$

where $n$ is an integer, $D$ is the exchange spin stiffness constant, $\mu_B$ is the Bohr magneton, $g$ is the g-factor, and $L$ is the sample thickness

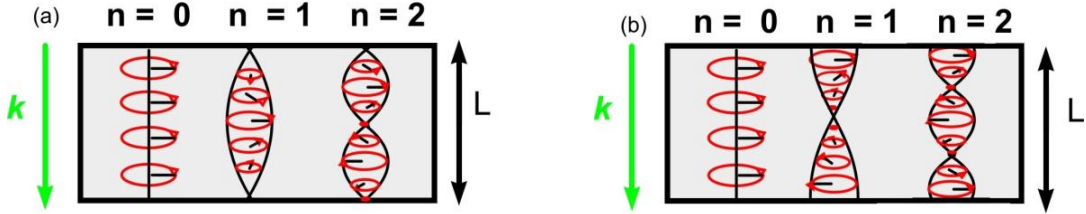

Fig. 12. Spin wave resonances in homogeneous thin magnetic films with a thickness $L$ that have a node (a) or maximum (b) at the surface; $n$ is the mode number.

In FMR only the modes with odd $n$ are observed [11] and the corresponding resonant fields are smaller than that of the uniform magnetization precession (i.e., $\Delta H_n > 0$). In the magneto-optical pump-and-probe experiment, the external magnetic field is kept constant during the measurement of any dynamical MO trace. Consequently, the SWRs are apparent as additional frequencies that are larger than that of the uniform magnetization precession. Ultrafast optical pulses also excite all resonant modes without any $k$ selectivity [18, 30]. Consequently, for a homogeneous magnetic film with a given thickness, a higher number of SWRs is detectable in the MO dynamical traces than in the FMR spectra. This is particularly important for (Ga,Mn)As that is magnetically homogeneous only when prepared in a form of

rather thin films and, therefore, where only a limited number of SWRs is present within a detectable range of the precession frequencies. For an external magnetic field $H_{ext}$ applied in the sample plane, the angular frequency of the $n$-th SWR mode $f_n$ is given by [3, 31]

$$f_n = \frac{g\mu_B}{h} \sqrt{\left(H_{ext}\cos(\varphi-\varphi_H) - 2K_{out} + \frac{K_c(3+\cos 4\varphi)}{2} + 2K_u \sin^2\left(\varphi-\frac{\pi}{4}\right) + \Delta H_n\right) \times \\ \times \left(H_{ext}\cos(\varphi-\varphi_H) + 2K_c \cos 4\varphi - 2K_u \sin(2\varphi) + \Delta H_n\right)},$$

(19)

which enables to convert the experimentally measured frequency spacing of individual modes to the field differences $\Delta H_n$ from which the magnitude of the spin stiffness $D$ can be evaluated using Eq. (18) (see Fig. 5 in the main paper).

As we illustrate in the following chapter, the magnetic homogeneity of the investigated epilayer is absolutely essential for a correct determination of $D$ from the measured SWR spacing. Therefore, the experimental results obtained in samples that had been prepared by etching the original 48 nm thick (Ga,Mn)As epilayer down to the thickness 39, 29 and 15 nm are of fundamental importance. In Fig. 13 we show the corresponding FFT spectra of the measured oscillatory MO signal. Clearly, the frequency $f_0$ of the lowest SWR does not depend on the film thickness. This confirms that the lowest observed SWR really corresponds to the uniform precession of magnetization and, moreover, it proves that this film is magnetically homogeneous. Also the spacing $\Delta H_1$ shows the expected [see Eq. (2)] linear dependence on $n^2$ and $1/L^2$ (see Fig. 5 in the main paper) that enables a reliable determination of the value of $D$. In Fig. 14 we compare the experimental data for 48 nm and 15 nm thick epilayers from which the FFT spectra depicted in Fig. 13 were computed. Clearly, the etching of the sample from 48 nm to 15 nm not only suppressed the higher SWRs, which is apparent from the purely sinusoidal shape of the data for the 15 nm film, but it also increased the precession damping, which is probably a consequence of a slight variation of the etched film thickness within the laser spot size of 25 μm. These data illustrate that the magneto-optical pump-and-probe experiment is a very sensitive diagnostic tool not only of the magnetic but also of the structural quality of thin magnetic films.

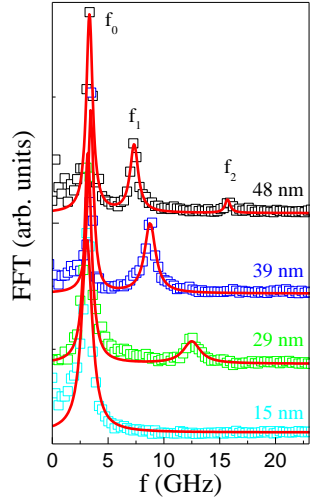

Fig. 13. Fourier spectra of oscillatory MO signals measured for $\mu_0 H_{ext}$ = 20 mT applied along the [010] crystallographic direction in samples prepared by etching from 48 nm thick (Ga,Mn)As epilayer with $x$ = 7% (points), the curves are labeled by the film thicknesses, normalized and vertically shifted for clarity; the lines are fits by a sum of Lorentzian peaks.

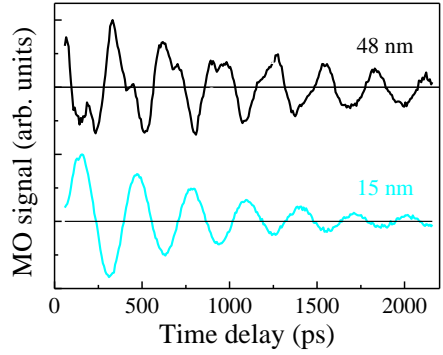

Fig. 14. Comparison of oscillatory parts of MO signals measured in the original 48 nm thick epilayer and in the epilayer that was etched down to 15 nm; the curves are normalized and vertically shifted for clarity. Experimental conditions are described in Fig. 13.

## DEMONSTRATION OF INAPPLICABILITY OF SPIN STIFFNESS MEASUREMENT IN THICK (GA,MN)AS EPILAYERS

Finally, we illustrate the significance of the film magnetic homogeneity for a correct evaluation of the spin stiffness. For this purpose we selected a 500 nm thick (Ga,Mn)As epilayer with 7% Mn (i.e, a sample with the same nominal Mn doping as the one used in experiments depicted in Fig. 13 and Fig. 14). In Fig. 15 we show the temperature dependent magnetization projections to several crystallographic directions measured in the as-grown and annealed samples. In the as-grown sample, the temperature dependence of magnetization projections is strongly non-monotonous [see Fig. 15(a)]. Moreover, the Curie temperature $T_c$

is only ≈ 60 K that is very low for a material with 7% Mn. This is a consequence of a high concentration of unintentional interstitial Mn impurities in the sample that compensate both the local moment and the holes produced by substitutional Mn atoms [32]. The amount of interstitial Mn impurities in the sample can be reduced by a thermal annealing [33]. However, even very long annealing times are not sufficient for obtaining a high quality sample from the thick epilayer due to the formation of the surface oxide that controls the outdiffusion of interstitial Mn impurities [33]. Therefore, the 40 h long annealing at 200 °C led to an increase of $T_c$ but only to 90 K, which is still substantially lower than $T_c \approx 150$ K observed in thin samples with the same nominal concentration of Mn. Simultaneously, the temperature dependence of magnetization does not show the expected sharply vanishing magnetization at $T_c$ (cf. Fig. 2 in the main paper for the data in optimized epilayers).

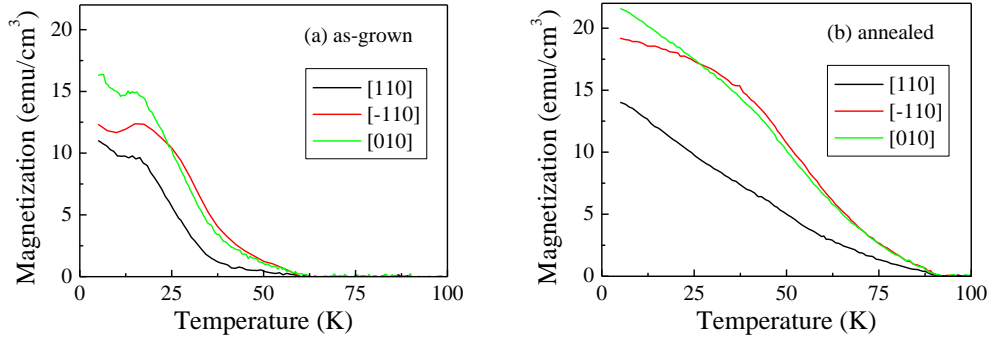

Fig. 15. Temperature dependence of the magnetization projections to different crystallographic directions measured by SQUID in 500 nm thick (Ga,Mn)As epilayer with $x = 7\%$. (a) As-grown sample. (b) Sample annealed for 40 hours at 200°C.

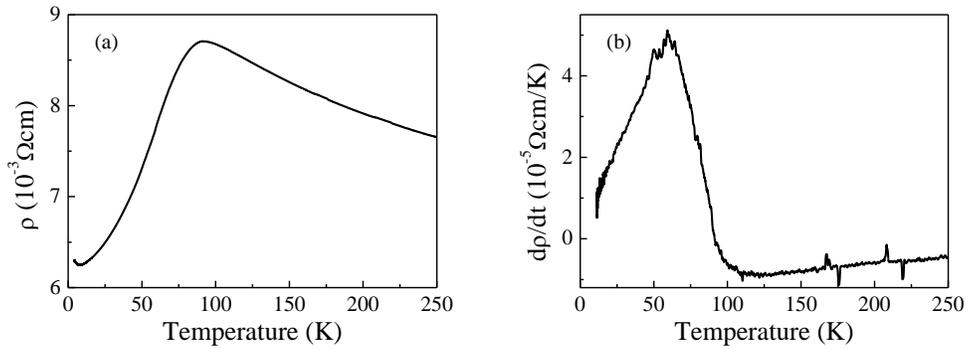

Fig. 16. Temperature dependence of the resistivity $\rho$ (a) and its temperature derivative $d\rho/dT$ (b) measured in 500 nm thick (Ga,Mn)As epilayer with $x = 7\%$ annealed for 40 hours at 200°C.

In Fig. 16 we show the temperature dependence of the resistivity and its temperature derivative measured in the annealed sample. Clearly, there is no sharp Curie point singularity

in the temperature derivative of the resistivity which is the fingerprint of a high magnetic quality of (Ga,Mn)As epilayer (cf. Fig. 1(a) in the main paper)

In Fig. 17 we show the time-resolved magneto-optical signals measured in this 500 nm thick epilayer. In the as-grown sample two precession modes can be identified. In the annealed sample the improved magnetic quality leads to a strong suppression of the magnetization precession damping with respect to that observed in the as-grown sample. For example, the data shown in Fig. 17(a) and (b) for the lowest modes correspond to damping times of 210 ps and 460 ps for the as-grown and annealed sample, respectively. In addition, the annealing led to a considerable increase of the number of observed SWR modes in the measured TRMO signal. However, their identification is a rather complicated task.

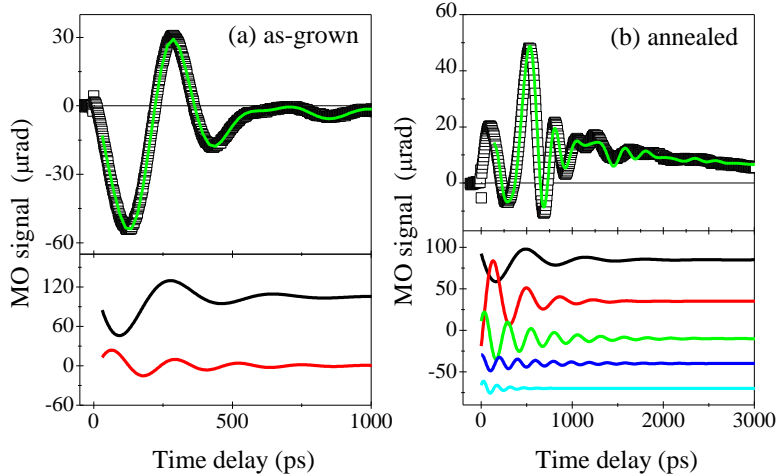

Fig. 17. Time-resolved magneto-optical signals (points) measured in as-grown (a) and annealed (b) 500 nm thick (Ga,Mn)As epilayer with $x = 7\%$; note the different $x$-scales in (a) and (b). The lines in the upper parts of the figures are a sum of damped harmonic functions and the corresponding precession modes are plotted in the lower parts of the figures. External magnetic field of 10 mT was applied along the [010] crystallographic direction.

In Fig. 18 we show the FFT spectrum of the oscillatory MO signals measured in the annealed sample for external magnetic fields of 10 mT and 20 mT. Even though the magnetic field change was rather small, the FFT spectra were changed dramatically. In particular, at 10 mT there are 3 peaks with comparable intensities (and 5 peaks in total) while at 20 mT there is only 1 strong peak (and 4 peaks in total). In Fig. 19 we show the dependence of the frequency of SWR modes on the external magnetic field – at least for the first sight, it is not apparent how to assign the observed modes to mode numbers defined by Eq. (18), which is the basic requirement for an evaluation of the spin stiffness from the measured data.

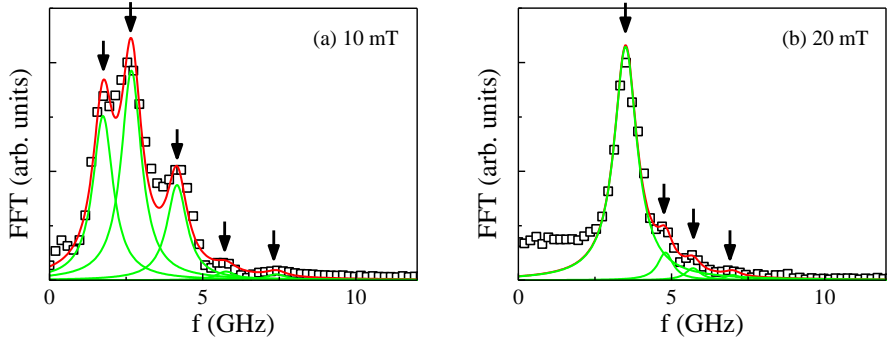

Fig. 18. Fourier spectrum of the oscillatory part of the MO signal measured in the annealed sample for external magnetic fields of 10 mT (a) and 20 mT (b) applied along the [010] crystallographic direction (points); the red line is a fit by a sum of Lorentzian peaks (green lines) and the arrows indicate positions of the peak frequencies.

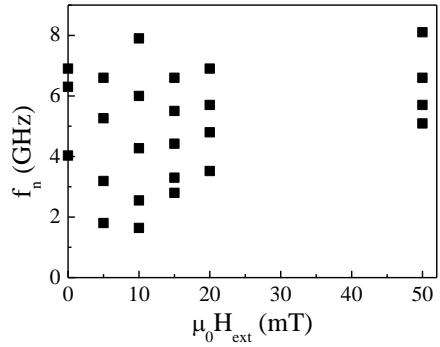

Fig. 19. Dependence of the precession frequency $f_n$ on $H_{ext}$ measured in the annealed sample for external magnetic field applied along the [010] crystallographic direction.

In Fig. 20 (a) we show a plausible assignment of the measured frequencies to four SWRs described in the previous chapter and one non-propagating surface mode [11]. We note that the identification of the lowest mode for fields below 15 mT as the surface mode is based on the analysis reported in Ref. 11 – in particular, due to the observations that this mode is apparent only at certain external magnetic fields and that it has a smaller amplitude that the one assigned to the homogeneous precession [see Fig. 18(a)]. Following the analysis reported in the previous chapter, we can now proceed to the evaluation of the spin stiffness. In Fig. 20 (b) the deduced values of $\Delta H_n$ are plotted as a function of $n^2$. The observed mode spacing deviates significantly from that expected for SWRs in a magnetically homogeneous film [see Eq. (18)] which is another fingerprint of the magnetic inhomogeneity in this 500 nm thick epilayer [9-12]. Consequently, despite a large number of SWRs detected in this sample, they cannot be used for a direct determination of the spin stiffness.

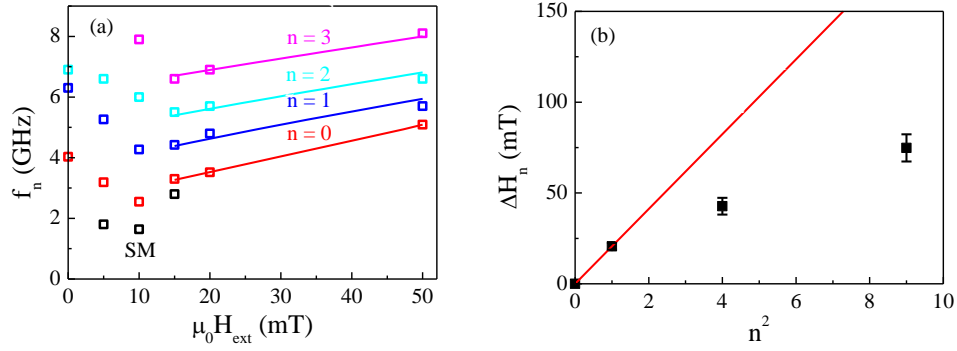

Fig. 20. (a) Dependence of the precession frequency $f_n$ on $H_{ext}$ measured in the annealed sample for external magnetic field applied along the [010] crystallographic direction re-plotted from Fig. 19 with the depicted assignment of precession frequencies to the individual SWRs and to the surface mode, SM (points). Lines are fits by Eq. (19). (b) Dependence of mode spacing $\Delta H_n$ on square of the mode number $n$ (points), line is the theoretical dependence $\Delta H_n \sim n^2$.

[9] Rappoport, T. G. *et al.* Anomalous behavior of spin-wave resonances in $Ga_{1-x}Mn_xAs$ thin films. Phys. Rev. B **69**, 125213 (2004).

[10] Zhou, Y., Cho, Y., Ge, Z., Liu, X., Dobrowolska, M., and Furdyna, J. K.

Magnetic Anisotropy, Spin Pinning, and Exchange Constants of (Ga,Mn)As Films.

*IEEE Trans. Magn.* **43**, 3019 (2007).

[11] Liu, X., Zhou, Y. Y., and Furdyna, J. K. Angular dependence of spin-wave resonances and surface spin pinning in ferromagnetic (Ga,Mn) As films. *Phys. Rev. B* **75**, 195220 (2007).

[12] Bihler, C., Schloch, W., Limmer, W., Goennenwein, S. T. B., and Brandt, M. S.

Spin-wave resonances and surface spin pinning in $Ga_{1–x}Mn_xAs$ thin films.

*Phys. Rev. B* **79**, 045205 (2009).

[13] Wang, D. M., Ren, Y. H., Liu, X., Furdyna, J. K., Grimsditch, M., and Merlin, R.

Light-induced magnetic precession in (Ga,Mn)As slabs: Hybrid standing-wave Damon-Eshbach modes. *Phys. Rev. B* **75**, 233308 (2007).

[14] Werpachowska, A., Dietl, T. Theory of spin waves in ferromagnetic (Ga,Mn)As.

*Phys. Rev. B* **82**, 085204 (2010).

[15] Jungwirth, T. *et al.* Prospects for high temperature ferromagnetism in (Ga,Mn)As semiconductors. *Phys. Rev. B* **72**, 165204 (2005).

[16] P. Němec, E. Rozkotová, N. Tesařová, F. Trojánek, E. De Ranieri, K. Olejník, J. Zemen, V. Novák, M. Cukr, P. Malý, T. Jungwirth, Experimental observation of the optical spin transfer torque. *Nat. Phys.* **8**, 411-415 (2012), arXiv: 1201.1436v1 *and its Supplementary material.*

[17] Tesařová, N., Němec, P., Rozkotová, E., Zemen, J., Trojánek, F., Olejník, K., Novák, V., Malý, P., and Jungwirth, T.

Experimental observation of the optical spin-orbit torque. submitted.

[18] van Kampen, M. *et al.* All-Optical Probe of Coherent Spin Waves.

*Phys. Rev. Lett.* **88**, 227201 (2002).

[19] Qi, J. *et al*. Coherent magnetization precession in GaMnAs induced by ultrafast optical excitation*. Appl. Phys. Lett*. **91**, 112506 (2007).

[20] Hashimoto, Y., Kobayashi, S. & Munekata, H. Photoinduced precession of magnetization in ferromagnetic (Ga,Mn)As. *Phys. Rev. Lett.* **100**, 067202 (2008).

[21] Hashimoto, Y. & Munekata, H. Coherent manipulation of magnetization precession in ferromagnetic semiconductor (Ga,Mn)As with successive optical pumping. *Appl. Phys. Lett.* **93**, 202506 (2008).
23



\end{document}